# The anisotropic grain size effect on the mechanical response of polycrystals: The role of columnar grain morphology in additively manufactured metals

S. Amir H. Motaman*, Dilay Kibaroglu

*Institut für Eisenhüttenkunde, RWTH Aachen University, Intzestraße 1, 52072 Aachen, Germany*

**ABSTRACT**

Additively manufactured (AM) metals exhibit highly complex microstructures, particularly in terms of grain morphology which typically features heterogeneous grain size distribution, irregular and anisotropic grain shapes, and the so-called columnar grains. The conventional morphological descriptors based on grain shape idealization are generally inadequate for representing complex and anisotropic grain morphology of AM microstructures. The primary aspect of microstructural grain morphology is the state of grain boundary spacing or grain size whose effect on the mechanical response is known to be crucial. In this paper, we formally introduce the notion of axial grain size from which we derive mean axial grain size, effective grain size, and grain size anisotropy as robust morphological descriptors capable of effectively representing highly complex grain morphologies. We instantiated a discrete sample of polycrystalline aggregate as a representative volume element (RVE) featuring random crystallographic orientation and misorientation distributions. However, the instantiated RVE incorporates the typical morphological features of AM microstructures including distinctive grain size heterogeneity and anisotropic grain size owing to its pronounced columnar grain morphology. We ensured that any anisotropy observed in the macroscopic mechanical response of the instantiated sample primarily originates from its underlying anisotropic grain size. The RVE was then employed for meso-scale full-field crystal plasticity simulations corresponding to uniaxial tensile deformation along various axes via a spectral solver and a physics-based crystal plasticity constitutive model which was developed, calibrated, and validated in earlier studies. Through the numerical analyses, we isolated the contribution of anisotropic grain size to the anisotropy in the mechanical response of polycrystalline aggregates, particularly those with the characteristic complex grain morphology of AM metals. This contribution can be described by an inverse square relation.

## 1. Introduction

The deformation behavior of polycrystalline materials is one of the most complex phenomena in Materials Science. The complexity of polycrystal deformation is attributed to the intricate hierarchical polycrystalline microstructure (i.e., the intrinsic multi-scale structure of polycrystalline materials) and its complex evolution during elasto-plastic deformation. One of the most complex aspects of polycrystal deformation behavior is the anisotropy in the macroscopic mechanical response, which arises due to the presence of crystallographic and/or morphological textures [1].

Metal additive manufacturing (AM) enables production of metallic polycrystals with extremely higher microstructural complexity compared to their conventional counterparts [2]. One of the characteristic complexities inherent in polycrystalline microstructures is associated with their grain morphology. The typical grain morphology in AM microstructures consists of heterogeneous multi-modal grain size distribution, anomalous and anisotropic grain shapes, and the so-called columnar grains which are ascribed to directional solidification and epitaxial grain growth during additive manufacturing [3–10].

---

* Corresponding author.
   E-mail address: seyedamirhossein.motaman@rwth-aachen.de (S.A.H. Motaman).

The crystalline microstructural texture/polarity can be classified into two categories: crystallographic texture and morphological texture [1,2]. Conventionally, morphological texture is represented using morphological descriptors associated with idealized grain shapes [11–14]. In this approach, the morphological descriptors are parameters related to the equivalent or best-fit spheres and ellipsoids of the grains within the microstructure: equivalent size, equivalent minimum aspect ratio, equivalent maximum aspect ratio, and parameters corresponding to the orientation of equivalent axes (Appendix A). The conventional morphological representation based on grain shape simplification presents two major shortcomings with respect to the heterogeneous microstructures processed by unconventional manufacturing technologies like metal additive manufacturing processes:

- Due to the multitude of morphological descriptors, constructing a distribution function that preserves the correlations between them is computationally demanding. Furthermore, a multi-variate joint probability distribution of grain morphology is intractable in terms of representation and analysis. Therefore, almost always, the distributions of these descriptors are treated without taking their correlations into account. In other words, it is presumed that the correlations between the conventional morphological descriptors are entirely random. Conversely, in AM metals, these descriptors often exhibit strong correlations. For example, the columnar grains commonly found in AM metals are characterized by their large equivalent size and highly elongated shapes (low equivalent minimum aspect ratio) whose equivalent major axes are predominantly oriented at low angles relative to the build direction (BD). In addition, the typical multi-modal distributions of the conventional morphological descriptors in AM microstructures are attributed to the aforementioned correlations.

- The grain shape idealization method is not often suitable to adequately represent the irregular and complex grain shapes found in unconventional microstructures, such as those typically observed in the microstructures of AM metals. For instance, ellipsoid simplification cannot sufficiently capture the anisotropy of columnar grains in AM microstructures [2]. As another example, in AM microstructures, numerous grains exhibit significant concavity, while sphere and ellipsoid are strictly convex. Additionally, in AM microstructures, it is not uncommon to encounter small grains completely enclosed by larger grains (i.e., having only one contiguous neighbor).

The most important aspect of microstructural grain morphology is the state of grain boundary spacing or grain size whose effect on the mechanical response is known to be significant. The effect of grain size on the yield stress of crystalline materials, which is known as the Hall-Petch effect, is one of the most investigated microstructural effects [15–25]. The grain boundary spacing also affects the strain hardening response of crystalline materials through its influence on confining the mean free path of dislocations [25–32]. In addition, the grain size plays an essential role in activating and suppressing different mechanisms accommodating crystal plasticity including twinning [33–37], stress/strain-induced phase transformation [38,39], and stress-induced/shear-coupled grain boundary motion (migration, sliding, and rotation) [40–46], as well as some macroscopically observable phenomena such as the inverse Hall-Petch effect in ultrafine-grained polycrystals [47–55], and yield point phenomenon or discontinuous yielding in fine-grained polycrystals [16,56–65].

Given the substantial effect of grain size on the mechanical response of crystalline materials, it is rational to define a single-variable morphological descriptor based on grain boundary spacing to effectively represent both the grain size and the morphological texture. The equivalent size of a grain approximates its average grain boundary spacing but it does not convey any information regarding the grain's anisotropic shape or spatial orientation (Appendix A). Even though the equivalent sphere/circle diameter is becoming the method of choice to approximate the (scalar) effective grain size, the mean of the so-called chord/intercept length distribution offers a more precise estimation of the effective grain size, particularly in microstructures with anisotropic grain shapes [66–68]. In addition, the so-called angularly resolved chord length distribution has been used to simultaneously represent the state of grain size and morphological texture of two-dimensional (2D) sections of polycrystalline microstructures [69], the state of anisotropic spacing of dispersed particles in 2D sections of a matrix, as well as the size and morphological texture of pores within a three-dimensional (3D) matrix [70]. In this context, the chord, or what we refer to as the axial grain size, is defined as a line segment parallel to a certain axis/direction whose end points lie at grain boundaries while its interior points are all located within a single grain. As such, the definition of the axial grain size or chord is not ambiguous. Nevertheless, given a microstructure and a probe axis, the estimation of the underlying axial grain size distribution (or chord length distribution) has not yet been properly conceptualized. The standard approach is that grids of scan/probe lines parallel to some axes/directions are intercepted with the microstructure to estimate the corresponding axial grain size distribution.

The equivalent grain size distribution in a microstructure can be represented using the number or volume/area-weighted distributions (Appendix A). In case of a heterogeneous (multi-modal) grain size distribution, such as that one typically observes in the microstructures of welded and AM metals, it is recommended to estimate the effective grain size using the volume-weighted distribution of equivalent grain size instead of the number-weighted distribution [1,71–74]. Particularly, in the microstructures where there are a small number of large grains with significant contributions to the aggregate's volume, the heterogeneity and bimodal character of the grain size distribution will be lost or highly underestimated in representation by number-weighted equivalent grain size distribution. Hence, the effect of the small number of large grains in the microstructure will not be properly expressed in the estimated effective grain size obtained from the corresponding number-weighted equivalent grain size distribution. Therefore, the most precise approach to simultaneously represent the state of grain size and morphological texture of heterogeneous microstructures with anisotropic and anomalous grain shapes is through appropriately weighted axial grain size distribution.

In this paper, first, we present a rigorous continuum theory for the axial grain size, the axial grain size distribution, and the grain size anisotropy. Then, we apply the theory to an instantiated discrete sample of (single-phase) polycrystalline aggregate which has the typical morphological features of AM metallic microstructures including distinctive size heterogeneity and morphological texture owing to its pronounced columnar grain morphology. Based on the notion of axial grain size, the columnar grains are characterized by their large axial grain size along BD relative to the axes orthogonal to BD. In the present study, we seek the contribution of the morphological texture and anisotropic grain size to the anisotropic mechanical response of polycrystals, particularly those with pronounced columnar grain morphology. Therefore, we isolated the morphological texture of the instantiated aggregate by assigning crystallographic orientation to its grains in such a way to obtain random crystallographic orientation and misorientation distributions (i.e., crystallographically textureless microstructure). As a result, we assume that any anisotropy in the macroscopic mechanical response must be associated with the underlying anisotropic grain size of the microstructure. The strain hardening anisotropy associated with the generated aggregate, which describes the anisotropy in its macroscopic strain hardening response under uniaxial deformation, was evaluated. To that end, the instantiated mesostructure sample was used as a representative volume element (RVE) for meso-scale full-field crystal plasticity simulations corresponding to uniaxial (tensile) deformation along different axes via a spectral solver and a physics-based crystal plasticity constitutive model which was developed, calibrated, and validated in previous studies [1,2].

## 2. Theory

To seek the contribution of anisotropic grain size to strain hardening anisotropy, first we need a continuum theory for axial grain size and grain size anisotropy as well as uniaxial strain hardening and strain hardening anisotropy. The grain size and the strain hardening are elements of microstructural and micromechanical states (i.e., micro-states), respectively. Even though the grain size anisotropy and the strain hardening anisotropy depends on those micro-states, they represent aspects of the macro-state (i.e., macroscopic state) of the material. In this section, we present formal definitions and theoretical background related to the aforementioned notions. In the following sections, in pursue of a relationship between the grain size and strain hardening anisotropies, we use this theoretical framework to calculate axial grain size, grain size anisotropy, uniaxial strain hardening, and strain hardening anisotropy of an instantiated discrete polycrystalline microstructure.

### 2.1. Microstructural grains and grain boundaries

The polycrystalline microstructure domain as a continuum medium can be divided into finite volumetric regions (subsets of microstructure domain) known as grains/crystallites. The grain boundary corresponding to a grain in a microstructure can be idealized as a closed topological surface embedded in Euclidean 3-space enclosing the grain interior, where each point on the grain boundary is associated with a relatively sharp transition (jump) of microstructural state (combination of crystallographic orientation and phase) across it beyond a certain threshold. Let the finite microstructure domain at time $t$, $\Omega \equiv \Omega(t) \subset \mathbb{R}^3$, be discretized to $N_G \in \mathbb{N}$ grains occupying the regions $g_1, g_2, \ldots, g_{N_G}$, and their grain boundaries $\partial g_1, \partial g_2, \ldots, \partial g_{N_G}$, so that:

$$\mathrm{G} \equiv \{g_i \subseteq \Omega \mid i \in \mathbb{N}; \ i \leq N_G\}; \quad \forall g_i, g_j \in \mathrm{G}, g_i \neq g_j : g_i \cap g_j = \emptyset;$$

$$\partial \mathrm{G} \equiv \{\partial g_i \subseteq \Omega \mid i \in \mathbb{N}; \ i \leq N_G\}; \quad \forall \partial g_i, \partial g_j \in \partial \mathrm{G}: \partial g_i \cap \partial g_j \begin{cases} \neq \emptyset & : g_i \text{ and } g_j \text{ are contiguous} \\ = \emptyset & : \text{otherwise} \end{cases};$$

$$\Omega = \bigcup_{i=1}^{N_G} g_i \cup \partial g_i = \mathrm{G} \cup \partial \mathrm{G}; \tag{1}$$

where $\mathrm{G}$ and $\partial \mathrm{G}$, respectively, represent the set of all grains and grain boundaries in the microstructure domain $\Omega$ which is simply connected; and $\emptyset \equiv \{\}$ denotes the empty set. The elements of $\mathrm{G}$ (grains) are pairwise disjoint and form a partition of $\Omega - \partial \mathrm{G}$. Moreover, according to the definition, two grains are contiguous if and only if the intersection of their boundaries is not the empty set.

### 2.2. Directions and axes

In Euclidean 3-space, a (3D) direction can be defined by a unit vector $\hat{\boldsymbol{r}}$ ($\|\hat{\boldsymbol{r}}\| = 1$). Let the direction $\hat{\boldsymbol{r}}$ be represented by $\hat{\boldsymbol{r}} \equiv (r_1, r_2, r_3) \equiv r_1 \hat{\boldsymbol{e}}_1 + r_2 \hat{\boldsymbol{e}}_2 + r_3 \hat{\boldsymbol{e}}_3$ in the right-handed Cartesian coordinate system $x_1 x_2 x_3$ with the orthonormal basis $\{\hat{\boldsymbol{e}}_1, \hat{\boldsymbol{e}}_2, \hat{\boldsymbol{e}}_3\}$, and by $\hat{\boldsymbol{r}} \equiv (\theta, \varphi) \equiv \theta \hat{\boldsymbol{e}}_\theta + \varphi \hat{\boldsymbol{e}}_\varphi + \hat{\boldsymbol{e}}_r$ in the spherical coordinate system $\theta \varphi r$ having the orthonormal basis $\{\hat{\boldsymbol{e}}_\theta, \hat{\boldsymbol{e}}_\varphi, \hat{\boldsymbol{e}}_r\}$ and the same origin as $x_1 x_2 x_3$. Note that we use the notations of 3-tuple and 2-tuple to represent the variable components/coordinates of $\hat{\boldsymbol{r}}$, in the coordinate systems $x_1 x_2 x_3$ and $\theta \varphi r$, respectively. As shown in Fig. 1, the polar angle $\theta$ denotes the angle between $\hat{\boldsymbol{r}}$ and $\hat{\boldsymbol{e}}_3$, and the azimuthal angle $\varphi$ is the angle between $\hat{\boldsymbol{e}}_1$ and the projection of $\hat{\boldsymbol{r}}$ on the $x_1 x_2$ plane ($\boldsymbol{r}_{12} \equiv (r_1, r_2, 0)$). This leads to the following coordinate transformation:

$$r_1 = \sin\theta \, \cos\varphi; \quad r_2 = \sin\theta \, \sin\varphi; \quad r_3 = \cos\theta. \tag{2}$$

Therefore, the direction $\hat{\boldsymbol{r}}$ is a vector from the origin of the unit sphere (i.e., the sphere with unit radius) to a particular point on the surface of the unit sphere. Thus, an arbitrary direction $\hat{\boldsymbol{r}}$ belongs to the 3D direction space (set of all 3D directions) which can be represented by the surface of the unit sphere (i.e., 2-sphere) ($S^2$):

$$S^2 \equiv \{\hat{\boldsymbol{r}} \in \mathbb{R}^3 \mid \|\hat{\boldsymbol{r}}\| = 1\} \equiv \{\hat{\boldsymbol{r}} = (\theta, \varphi) \mid -\pi < \theta \leq \pi, 0 \leq \varphi < \pi\}; \tag{3}$$

where $\|\bullet\|$ denotes the Euclidean norm. Direction $\hat{r}$ as well as its inverse $(-\hat{r})$ uniquely define an axis. Hence, an axis is represented by $\hat{r}$ or $\pm\hat{r}$, while a line is defined by an axis and at least one point. Therefore, the surface of a unit hemisphere is a fundamental/asymmetrical region/zone of the axis space which is the set of all possible 3D axes. The fundamental region of the axis space has the property that there is an injective/one-to-one map between its distinct elements to distinct axes, and thus it corresponds to the fundamental region of direction space with antipodal symmetry, which is conventional to be represented by the surface of the "northern" unit hemisphere ($S_{z+}^2$):

$$S_{z+}^2 \equiv \left\{ \hat{r} = (\theta, \varphi) \in S^2 \,\Big|\, -\frac{\pi}{2} \le \theta < \frac{\pi}{2},\, 0 \le \varphi < \pi \right\}; \tag{4}$$

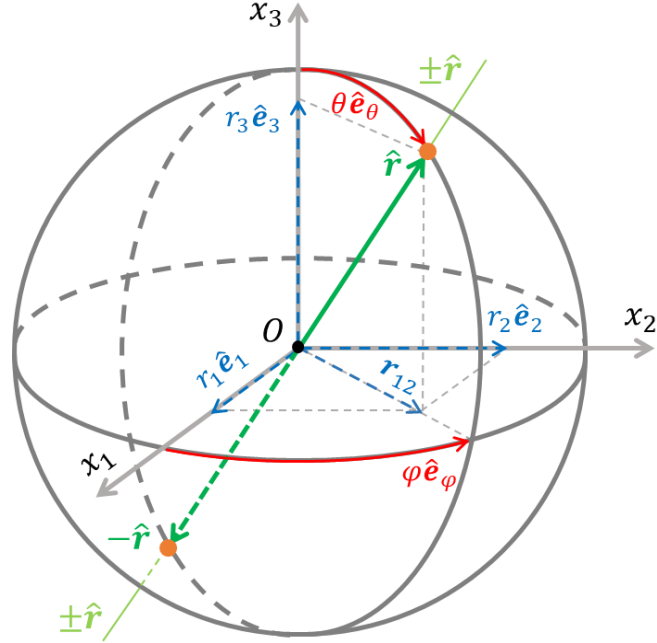

**Fig. 1.** Schematic illustration of 3D directions and axes, and their coordinates in Cartesian and spherical frames. Note that the radius of the sphere is unity (i.e., unit sphere).

### 2.3. Axial grain size

Let $\Omega \subset \mathbb{R}^3$ be the microstructure domain at time $t$. The axial grain size $D(\hat{r}, x)$ associated with the axis $\hat{r}$ at point $x$ with the position (vector) $x = (x_1, x_2, x_3)$ (represented in the $x_1 x_2 x_3$ frame) in the grain $g_i$ ($x \in g_i$) within the microstructure domain $\Omega$ ($g_i \subseteq \Omega$) is the length of the unique line segment $\overline{ab}$ which (i) is parallel to $\hat{r}$ ($\overline{ab} \parallel \hat{r}$), (ii) is fully located in $g_i$ ($\overline{ab} \subset g_i$), (iii) includes $x$ ($x \in \overline{ab}$), and (iv) its endpoints ($a$ and $b$) are located at the grain boundary of $g_i$ ($\{a, b\} \subset \partial g_i$):

$$D(\hat{r}, x) \equiv \|\overline{ab}\| \equiv \|b - a\| \equiv \sqrt{(b_1 - a_1)^2 + (b_2 - a_2)^2 + (b_3 - a_3)^2}; \tag{5}$$

where $a = (a_1, a_2, a_3)$ and $b = (b_1, b_2, b_3)$ represent the coordinates of $a$ and $b$ with respect to the Cartesian frame $x_1 x_2 x_3$, respectively. The axial grain size $D(\hat{r}, x)$ is schematically shown in Fig. 2. Since both directions $\hat{r}$ and $-\hat{r}$ define one unique axis, the axial grain size has the antipodal symmetry ($D(\hat{r}, x) \equiv D(-\hat{r}, x)$), and hence its fundamental domain corresponds to the fundamental domain of direction space with antipodal symmetry ($S_{z+}^2$).

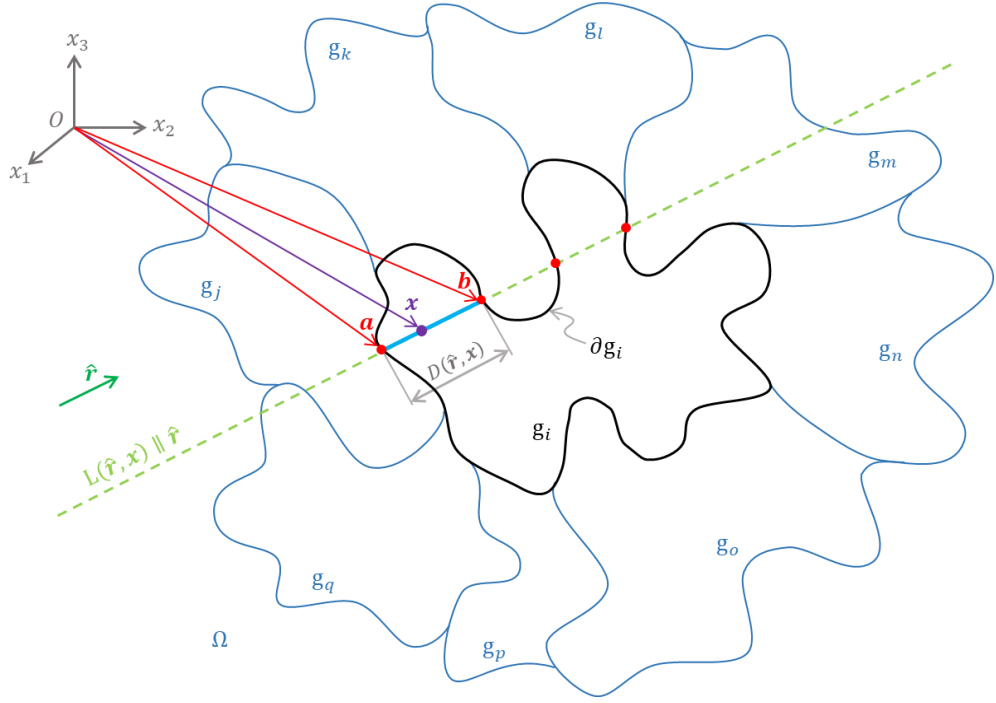

**Fig. 2.** Schematic illustration of axial grain size associated with the axis $\hat{r}$ at the point $x$ in a 2D section of a hypothetical microstructure domain $\Omega$.

According to Eq. (5), measuring the axial grain size associated with the axis $\hat{r}$ at point $x$ ($D(\hat{r}, x)$) is reduced to finding the positions of points $a$ and $b$. To find $a$ and $b$, first, the line $L(\hat{r}, x)$ parallel to $\hat{r}$ passing through $x$ must be identified:

$$L(\hat{r}, x) = \{(l_1, l_2, l_3) \mid l_1 = r_1 s + x_1, \; l_2 = r_2 s + x_2, \; l_3 = r_3 s + x_3, \; s \in \mathbb{R}\}; \quad x \in \Omega; \quad \hat{r} \in S^2_{z+}; \quad (6)$$

where $s$ is the line parameter. Based on the line parametrization in Eq. (6), at $s = 0$: $(l_1, l_2, l_3) = (x_1, x_2, x_3)$. Points $a$ and $b$ belong to the intersection of line $L(\hat{r}, x)$ and the grain boundary $\partial g_i$ ($\{a, b\} \subset L(\hat{r}, x) \cap \partial g_i$). In fact, $a$ and $b$ are the nearest points in $L(\hat{r}, x) \cap \partial g_i$ to $x$, while they are on different sides of $x$ along $L$ as $x \in \overline{ab}$. Therefore, given the line parametrization in Eq. (6), the sign of the line parameter $s$ is different for $a$ and $b$ ($s|_a \, s|_b < 0$). Consequently, the intersection set $\Psi(L, \partial g_i)$, which contains the line parameters corresponding to the intersection points of the grain boundary $\partial g_i$ and the line $L(\hat{r}, x)$, is given as:

$$\Psi(L, \partial g_i) = \left\{ s = \frac{l_1^{(i)} - x_1}{r_1} = \frac{l_2^{(i)} - x_2}{r_2} = \frac{l_3^{(i)} - x_3}{r_3} \,\middle|\, (l_1^{(i)}, l_2^{(i)}, l_3^{(i)}) \in L(\hat{r}, p) \cap \partial g_i \right\}. \quad (7)$$

Finally, the coordinates of $a$ and $b$ in the Cartesian frame $x_1 x_2 x_3$ are calculated as follows:

$$\begin{cases} a_1 = r_1 s_a + x_1 \\ a_2 = r_2 s_a + x_2 \,; \\ a_3 = r_3 s_a + x_3 \end{cases} s_a = \max\{s \in \Psi(L, \partial g_i) \mid s < 0\};$$

$$\begin{cases} b_1 = r_1 s_b + x_1 \\ b_2 = r_2 s_b + x_2 \,; \\ b_3 = r_3 s_b + x_3 \end{cases} s_b = \min\{s \in \Psi(L, \partial g_i) \mid s > 0\}; \quad (8)$$

By inserting Eq. (8) into Eq. (5), and considering the normalization condition $\|\hat{r}\| \equiv 1$, we arrive at:

$$D(\hat{r}, x) = |s_b - s_a|. \tag{9}$$

Moreover, the minimum distance to grain boundary along the axis $\hat{r}$ at point $x$ ($D_m(\hat{r}, x)$) is obtained as follows:

$$D_m(\hat{r}, x) \equiv \min\{|s| \mid s \in \Psi(L, \partial g_i)\} \equiv \min\{|s_a|, |s_b|\}. \tag{10}$$

*2.4. Grain size anisotropy*

The axial grain size distribution associated with the axis $\hat{r}$ ($P(D|\hat{r})$) is the conditional probability density of measuring the axial grain size $D(\hat{r}, x)$ at a random point $x$ in the microstructure given the axis $\hat{r}$. Intuitively, given a microstructure domain $\Omega \subset \mathbb{R}^3$ and axis $\hat{r} \in S^2$, one can think of $P(D|\hat{r})dD$ as the differential volume fraction of material points corresponding to axial grain size falling in an infinitesimal range around $D$:

$$P(D|\hat{r})dD \equiv \left.\frac{dv}{V}\right|_{D,\hat{r}}; \quad V \equiv \int_\Omega dv; \quad \int_{-\infty}^{\infty} P(D|\hat{r})dD \equiv 1; \tag{11}$$

where $V$ is the volume of the microstructure domain $\Omega$; $dv$ denotes a differential volume; and $\left.\frac{dv}{V}\right|_{D,\hat{r}}$ is the notation for the differential volume fraction of material points in $\Omega$, which are associated with axial grain size (given the axis $\hat{r}$) falling within the infinitesimal interval $\left[D - \frac{1}{2}dD,\ D + \frac{1}{2}dD\right]$. The second integral in Eq. (11) expresses the normalization property of the conditional probability density function $P(D|\hat{r})$. Therefore,

$$\overline{D}(\hat{r}) \equiv \frac{1}{V}\int_\Omega D(\hat{r}, x)d^3x \equiv \int_{-\infty}^{\infty} P(D|\hat{r})D dD; \tag{12}$$

where $d^3x \equiv dx_p dy_p dz_p$ denotes a differential volume in the vicinity of point $x$; and $\overline{D}(\hat{r})$ is the mean axial grain size associated with the axis $\hat{r}$ or the mean grain boundary spacing along the axis $\hat{r}$. Note that the mean axial grain size ($\overline{D}(\hat{r})$) and the axial grain size distribution ($P(D|\hat{r})$) functions inherit the antipodal symmetry of the axial grain size ($D(\hat{r}, x)$): $\overline{D}(\hat{r}) \equiv \overline{D}(-\hat{r})$ and $P(D|\hat{r}) \equiv P(D|-\hat{r})$.

Furthermore, the effective grain size ($\overline{\overline{D}}$) is formally defined as the mean of the volume-weighted distribution of axial grain sizes corresponding to all axes $\hat{r} \in S^2_{z+}$ and material points $x \in \Omega$:

$$\overline{\overline{D}} \equiv \frac{\int_{S^2_{z+}}\int_\Omega D(\hat{r}, x)d^3x\, d^2\hat{r}}{\int_{S^2_{z+}}\int_\Omega d^3x\, d^2\hat{r}} \equiv \frac{1}{2\pi}\int_{S^2_{z+}} \overline{D}(\hat{r})d^2\hat{r}; \tag{13}$$

Where $d^2\hat{r} \equiv \sin\theta\, d\theta d\varphi$ denotes a differential area on the surface of the unit sphere in the vicinity of the direction $\hat{r}$ (i.e., invariant measure on $S^2$); and $\int_{S^2_{z+}} d^2\hat{r} = 2\pi$ is the size of the fundamental region of axis space (i.e., the area of the unit-hemispherical surface). In other words, the effective grain size $\overline{\overline{D}}$ is the expectation value of the probability distribution $P(D)$ which returns the probability density of grain boundary spacing $D$ in the given microstructure:

$$\bar{\bar{D}} \equiv \int_{-\infty}^{\infty} P(D) D \mathrm{d}D\,; \quad P(D) \equiv \int_{S^2_{z+}} P_{Dr}(D,\hat{r}) \mathrm{d}^2\hat{r}\,; \quad \forall \hat{r} \in S^2_{z+}\colon P_{Dr}(D,\hat{r}) \equiv P(D|\hat{r}) P_r(\hat{r})\,; \qquad (14)$$

where $P_{Dr}(D,\hat{r})$ denotes the joint probability density of random variables $D$ and $\hat{r}$; and $P_r(\hat{r}) \equiv \frac{1}{2\pi}$ is the probability density of the axis $\hat{r}$ within the axis space. Combining Eqs. (12) and (14) leads to Eq. (13).

Given a microstructure and Eqs. (12) and (13), we define the grain size anisotropy ($\widehat{D}(\hat{r})$) as mean axial grain size normalized by the effective grain size:

$$\widehat{D}(\hat{r}) \equiv \frac{\bar{D}(\hat{r})}{\bar{\bar{D}}}\,; \quad \hat{r} \equiv (\theta, \varphi) \in S^2. \qquad (15)$$

The defined dimensionless grain size anisotropy function $\widehat{D}(\hat{r})$ is a map from the surface of the unit sphere to real numbers ($\widehat{D}\colon S^2 \to \mathbb{R}$) and has the unit of "multiple of random distribution" (mrd) representing the state of grain size anisotropy of the microstructure. According to the definitions (Eqs. (13) and (15)), the average of $\widehat{D}(\hat{r})$ over its domain is unity: $\frac{1}{4\pi}\int_{S^2} \widehat{D}(\hat{r}) \mathrm{d}^2\hat{r} \equiv \frac{1}{2\pi}\int_{S^2_{z+}} \widehat{D}(\hat{r}) \mathrm{d}^2\hat{r} \equiv 1$. Note that $\widehat{D}(\hat{r})$ inherits the antipodal symmetry of $\bar{D}(\hat{r})$ and $D(\hat{r},x)$: $\widehat{D}(\hat{r}) \equiv \widehat{D}(-\hat{r})$. Given an axis $\hat{r}$, $\widehat{D}(\hat{r}) > 1$ indicates that the mean axial grain size associated with $\hat{r}$ is greater than the effective grain size ($\bar{D}(\hat{r}) > \bar{\bar{D}}$); and if $\widehat{D}(\hat{r}) < 1$, the mean grain boundary spacing along $\hat{r}$ is lower than the effective grain size ($\bar{D}(\hat{r}) < \bar{\bar{D}}$). The function $\widehat{D}(\hat{r})$ is a fundamental morphological descriptor of the microstructure, which gives a representation of the state of the morphological texture. The function $\widehat{D}(\hat{r})$ and the scalar $\bar{\bar{D}}$ together provide a representation of grain morphology (combined morphological size and morphological texture). In that sense, the axial grain size renders a more comprehensive representation of microstructural grain morphology compared to the conventional representations of grain morphology via grain shape idealization (Appendix A).

Furthermore, one may apply the preceding definitions to the individual grains of a microstructure. Assuming a spherical grain with diameter $D_s$, we readily find that $\bar{\bar{D}} = \frac{2}{\pi} D_s$. This is very close to the stereological relationship proposed by [75]. It is intuitive that the mean axial grain size of a spherical grain is lower than its diameter ($\bar{\bar{D}} < D_s$) as the diameter of the sphere is the maximum possible axial grain size that can be measured in the sphere ($\forall \hat{r} \in S^2\colon 0 < \bar{D}(\hat{r}) \leq D_s$). Moreover, the mean axial grain size of a hypothetical spherically symmetric grain is independent of the probe axis and has a constant value for any arbitrary axis ($\forall \hat{r} \in S^2\colon \bar{D}(\hat{r}) = \bar{\bar{D}}$). Therefore, in a spherical grain: $\forall \hat{r} \in S^2\colon \widehat{D}(\hat{r}) = 1$, as sphere is the only perfectly isotropic shape. Since each continuous function $\widehat{D}(\hat{r})$ uniquely defines a 3D shape, each microstructure with the grain size anisotropy function $\widehat{D}(\hat{r})$ has a unique equivalent grain shape which is represented by $\widehat{D}(\hat{r})$.

The grain size anisotropy function $\widehat{D}(\hat{r})$ is a periodic real square-integrable antipodally symmetric function on the surface of the unit sphere ($S^2$). In other words, $\widehat{D} \in L^2_{\mathbb{R}\pm}(S^2)$, where $L^2_{\mathbb{R}\pm}(S^2)$ denotes the Hilbert space of real square-integrable functions with antipodal symmetry on the surface of the unit sphere (Appendix B). Therefore, the grain size anisotropy function $\widehat{D}(\hat{r})$ can be expressed by a series expansion of spherical harmonic basis functions (Appendix B) as follows:

$$\widehat{D}(\hat{r}) = \sum_{l=0,2,4}^{\infty} \sum_{m=0}^{l} \left( \widetilde{D}_l^m Y_l^m(\hat{r}) + (-1)^{-m} \left(\widetilde{D}_l^m\right)^* Y_l^{-m}(\hat{r}) \right); \qquad (16)$$

where $Y_l^m: S^2 \to \mathbb{C}$ denotes the spherical harmonic function (SHF) of degree $l \in \mathbb{N}_0$ and order $m \in \{m \in \mathbb{Z} \mid |m| \leq l\}$; $\widetilde{D}_l^m \in \mathbb{C}$ is the Fourier coefficient of SHF of degree $l$ and order $m$ associated with the function $\widehat{D}$; and asterisk $*$ indicates the complex conjugate. Since the SHFs form a complete set of orthonormal basis functions for the Hilbert space $L^2_{\mathbb{R}_\pm}(S^2)$, the Fourier coefficient $\widetilde{D}_l^m$ is the inner product of the SHF $Y_l^m$ with the function $\widehat{D} \in L^2_{\mathbb{R}_\pm}(S^2)$:

$$\widetilde{D}_l^m \equiv \langle Y_l^m | \widehat{D} \rangle \equiv \int_{S^2} \widehat{D}(\hat{\boldsymbol{r}}) \big(Y_l^m(\hat{\boldsymbol{r}})\big)^* \mathrm{d}^2 \hat{\boldsymbol{r}}. \tag{17}$$

Furthermore, given a microstructure, we define the grain size anisotropy index as a dimensionless scalar measure characterizing the average deviation from the fully isotropic counterpart of the microstructure:

$$\mathcal{A}[\widehat{D}] \equiv \sqrt{\langle \widehat{D} \rangle - 4\pi}; \tag{18}$$

where $\mathcal{A}$ denotes the anisotropy index functional; $\mathcal{A}[\widehat{D}] \geq 0$ is the grain size anisotropy index; $\langle f \rangle \in \mathbb{R}$ denotes the $L^2$-norm of function $f \in L^2(S^2)$ (Appendix B). As $\widehat{D} \in L^2_{\mathbb{R}}(S^2)$, the $L^2$-norm of $\widehat{D}$ reads $\langle \widehat{D} \rangle \equiv \int_{S^2} \big(\widehat{D}(\hat{\boldsymbol{r}})\big)^2 \mathrm{d}^2 \hat{\boldsymbol{r}}$. Since $\widehat{D}$ is properly normalized, based on the Cauchy-Schwarz inequality: $\langle \widehat{D} \rangle \geq 4\pi$. For a (hypothetical) microstructure having an isotropic grain size distribution: $\mathcal{A}[\widehat{D}] = 0$. Therefore, the isotropic case renders the lower bound of $\mathcal{A}[\widehat{D}]$, while there is no upper bound for the grain size anisotropy index.

## 2.5. Strain hardening anisotropy

Uniaxial strain hardening is a suitable measure to evaluate the anisotropic mechanical response. Firstly, it does not require finding a distinct border as the point of transition from macroscopic fully elastic deformation to macroscopic elasto-plastic deformation (i.e., macroscopic yield point) as treated in the phenomenological orthotropic yield functions [76–80]. Such a macroscopic transition does not occur at an instant during deformation and is often associated with a notably wide strain range, particularly in the absence of yield point elongation, where the arbitrary notion of "proof stress" is used to represent the macroscopic yield stress. Secondly, the term "anisotropy" indicates variation of a certain feature, property, or response of a material along different axes. Therefore, measuring the macroscopic mechanical response of the material to complex multiaxial macroscopic boundary conditions does not explicitly reveal the state of the anisotropy in the macroscopic mechanical response. Thirdly, at small strains, the anisotropy in the stress response of strongly textured polycrystalline materials might remain hidden [1,81]. However, since the strain hardening response is a derivative quantity (the derivative of the stress response with respect to strain), it is extremely sensitive to small details in microstructural texture and thus can unravel the anisotropy underlying the mechanical response.

The mean uniaxial strain hardening associated with the load axis $\hat{\boldsymbol{r}}$ within the fixed equivalent macroscopic true strain interval $[0, \varepsilon_f]$ ($\overline{H}(\hat{\boldsymbol{r}})$) is defined as follows:

$$\overline{H}(\hat{\boldsymbol{r}}) \equiv \frac{1}{\varepsilon_f} \int_0^{\varepsilon_f} H(\hat{\boldsymbol{r}}, \varepsilon) \mathrm{d}\varepsilon; \tag{19}$$

where $\varepsilon_f > 0$ is a constant final equivalent macroscopic true strain; and $H(\hat{\boldsymbol{r}}, \varepsilon)$ denotes the instantaneous equivalent macroscopic strain hardening (Eq. (C.9) in Appendix C) response of microstructure $\Omega$ subjected to the homogeneous boundary conditions corresponding to the uniaxial deformation along the (load) axis $\hat{\boldsymbol{r}}$ at equivalent macroscopic true strain $\varepsilon$, constant macroscopic nominal strain rate $\dot{\varepsilon}_0$, and constant temperature $T$. Given $\bar{H}$, one readily arrives at the instantaneous macroscopic stress response of the microstructure under uniaxial deformation along $\hat{r}$ at the final equivalent macroscopic true strain $\varepsilon_f$ by using $\bar{H}$ as linear operator: $\sigma \equiv \bar{H}\varepsilon_f$. Therefore, $\bar{H}(\hat{r})$ fully represents the macroscopic mechanical response of the microstructure subjected to uniaxial deformation at the constant macroscopic nominal strain rate $\dot{\varepsilon}_0$ and constant temperature $T$. Note that, analogous to the mean axial grain size ($\bar{D}$), the mean uniaxial strain hardening $\bar{H}$ is antipodally symmetric: $\bar{H}(\hat{r}) \equiv \bar{H}(-\hat{r})$. Furthermore, the effective uniaxial strain hardening ($\bar{\bar{H}}$) is defined as the volume-weighted average of mean uniaxial strain hardenings corresponding to all axes $\hat{r} \in S^2_{z+}$ and the equivalent macroscopic strain interval $[0, \varepsilon_f]$:

$$\bar{\bar{H}} \equiv \frac{\int_{S^2_{z+}} \int_0^{\varepsilon_f} H(\hat{r}, \varepsilon) \mathrm{d}\varepsilon \, \mathrm{d}^2\hat{r}}{\int_{S^2_{z+}} \int_0^{\varepsilon_f} \mathrm{d}\varepsilon \, \mathrm{d}^2\hat{r}} \equiv \frac{1}{2\pi} \int_{S^2_{z+}} \bar{H}(\hat{r}) \mathrm{d}^2\hat{r}. \tag{20}$$

Now, we can define the dimensionless strain hardening anisotropy ($\hat{H}(\hat{r})$) associated with the load axis $\hat{r}$ as the mean uniaxial strain hardening normalized by the effective uniaxial strain hardening:

$$\hat{H}(\hat{r}) \equiv \frac{\bar{H}(\hat{r})}{\bar{\bar{H}}}; \quad \hat{r} \equiv (\theta, \varphi) \in S^2. \tag{21}$$

Given the axis $\hat{r}$, $\hat{H}(\hat{r}) > 1$ indicates that the mean uniaxial strain hardening associated with the load axis $\hat{r}$ is greater than the effective uniaxial strain hardening ($\bar{H}(\hat{r}) > \bar{\bar{H}}$); and if $\hat{H}(\hat{r}) < 1$, the mean uniaxial strain hardening along the load axis $\hat{r}$ is lower than the effective uniaxial strain hardening ($\bar{H}(\hat{r}) < \bar{\bar{H}}$). Strain hardening anisotropy function $\hat{H}(\hat{r})$ is a map from the surface of the unit sphere to real numbers ($\hat{H}: S^2 \to \mathbb{R}$) and has the unit of mrd representing the anisotropy state in the uniaxial strain hardening of the microstructure. Due to the normalization, the average of $\hat{H}(\hat{r})$ over its doamin is unity: $\frac{1}{4\pi} \int_{S^2} \hat{H}(\hat{r}) \mathrm{d}^2\hat{r} \equiv \frac{1}{2\pi} \int_{S^2_{z+}} \hat{H}(\hat{r}) \mathrm{d}^2\hat{r} \equiv 1$. Note that $\hat{H}$ inherits the antipodal symmetry of $\bar{H}$: $\hat{H}(\hat{r}) \equiv \hat{H}(-\hat{r})$.

Analogous to the grain size anisotropy function $\hat{D}(\hat{r})$, the strain hardening anisotropy $\hat{H}(\hat{r})$ is a periodic real square-integrable antipodally symmetric function on the surface of the unit sphere ($\hat{H} \in L^2_{\mathbb{R}\pm}(S^2)$). Therefore, the strain hardening anisotropy function $\hat{H}(\hat{r})$ can be expressed by series expansion of spherical harmonic basis functions (Appendix B) as follows:

$$\hat{H}(\hat{r}) = \sum_{l=0,2,4}^{\infty} \sum_{m=0}^{l} \left( \widetilde{H}_l^m Y_l^m(\hat{r}) + (-1)^{-m} \left(\widetilde{H}_l^m\right)^* Y_l^{-m}(\hat{r}) \right); \tag{22}$$

where $\widetilde{H}_l^m \in \mathbb{C}$ is the Fourier coefficient of SHF of degree $l$ and order $m$ associated with the function $\hat{H}$. Since the SHFs form a complete set of orthonormal basis functions for the Hilbert space $L^2_{\mathbb{R}\pm}(S^2)$, the Fourier coefficient $\widetilde{H}_l^m$ is the inner product of the SHF $Y_l^m$ with the function $\hat{H}$:

$$\widetilde{H}_l^m \equiv \langle Y_l^m | \hat{H} \rangle \equiv \int_{S^2} \hat{H}(\hat{r}) \left(Y_l^m(\hat{r})\right)^* \mathrm{d}^2\hat{r}. \tag{23}$$

Furthermore, given a microstructure, we define the strain hardening anisotropy index as a dimensionless scalar measure indicating the mean deviation from its fully isotropic counterpart:

$$\mathcal{A}[\widehat{H}] \equiv \sqrt{\langle \widehat{H} \rangle - 4\pi}; \tag{24}$$

where $\mathcal{A}[\widehat{H}]$ is the strain hardening anisotropy index. Since $\widehat{H}$ is properly normalized, based on the Cauchy-Schwarz inequality: $\langle \widehat{H} \rangle \geq 4\pi$. For a microstructure with an isotropic strain hardening: $\mathcal{A}[\widehat{H}] = 0$.

We hypothesize that any macroscopic anisotropy in the initial microstructure $\Omega_0$ contributes to the overall anisotropy in the mean (macroscopic) uniaxial strain hardening. The sources of microstructural anisotropy in crystalline materials are crystallographic and morphological textures/polarities [2]. Assuming that, in a hypothetical crystalline microstructure, the only source of microstructural anisotropy is morphological texture, which is best represented by the grain size anisotropy function, one can isolate the contribution of the grain size anisotropy to the strain hardening anisotropy. According to Eqs. (19)-(21), to derive the strain hardening anisotropy function $\widehat{H}(\widehat{r})$, the initial microstructure $\Omega_0$ must be uniaxially deformed to the final equivalent macroscopic strain $\varepsilon = \varepsilon_f$ along different load axes. For large $\varepsilon_f$, the deformed microstructure $\Omega$ is appreciably different than its initial state $\Omega_0$ mainly due to plasticity. Nevertheless, the mean uniaxial strain hardening associated with any axis $\widehat{r}$ ($\overline{H}(\widehat{r})$) and thus the strain hardening anisotropy $\widehat{H}(\widehat{r})$ depend on the grain size anisotropy of the initial microstructure $\widehat{D}_0(\widehat{r})$ (i.e., the initial grain size anisotropy). Therefore, we seek the relationship between the functions $\widehat{H}(\widehat{r})$ and $\widehat{D}_0(\widehat{r})$, which can be expressed via functional $\mathcal{F}$ so that $\widehat{H} \equiv \mathcal{F}[\widehat{D}_0]$. In the following, for brevity we drop the subscript 0 from quantities associated with the axial grain size of the initial microstructure $\Omega_0$ as the axial and anisotropic grain size analysis is always carried out on the initial microstructure. Henceforth, the quantities $D$, $\overline{D}$, $\overline{\overline{D}}$, and $\widehat{D}$ all correspond to the initial domain $\Omega_0$. The notions introduced and defined in this section are summarized in Table 1.

**Table 1.** Summary of the introduced notions.

| | | |
|---|---|---|
| **Grain size** | $D(\widehat{r}, x)$ | Axial grain size at the material point $x$ along the axis $\widehat{r}$ |
| | $D_m(\widehat{r}, x)$ | Minimum distance to the nearest grain boundary at the material point $x$ along the axis $\widehat{r}$ |
| | $\overline{D}(\widehat{r})$ | Mean axial grain size along the axis $\widehat{r}$ |
| | $\widehat{D}(\widehat{r})$ | Grain size anisotropy of the axis $\widehat{r}$ |
| | $\overline{\overline{D}}$ | Effective grain size |
| **Strain hardening** | $H(\widehat{r}, \varepsilon)$ | Instantaneous uniaxial strain hardening associated with the load axis $\widehat{r}$ at the equivalent true strain $\varepsilon$ |
| | $\overline{H}(\widehat{r})$ | Mean uniaxial strain hardening associated with the load axis $\widehat{r}$ |
| | $\widehat{H}(\widehat{r})$ | Strain hardening anisotropy of the (load) axis $\widehat{r}$ |
| | $\overline{\overline{H}}$ | Effective strain hardening |

## 3. Methods

Given an initial microstructure $\Omega_0$, a point $x_0 \in \Omega_0$, and an axis $\widehat{r} \in S^2_{z+}$, one can unambiguously measure the axial grain size $D(x_0, \widehat{r})$. However, to obtain the axial grain size distribution $P(D|\widehat{r})$, a spatial discretization/binning/griding/meshing of the microstructure domain $\Omega_0$ is required. Nonetheless, the result of any experimental microstructure characterization, microstructure evolution simulation, or microstructure instantiation is always a discrete microstructure sample. Therefore, the measured axial grain size distribution is, in fact, an approximation of the underlying probability density function $P(D|\widehat{r})$. In order to obtain the grain size anisotropy $\widehat{D}(\widehat{r})$ and strain hardening anisotropy $\widehat{H}(\widehat{r})$, one also needs a discretization of the axis space $S^2_{z+}$ (i.e., binning of spherical coordinates $(\theta, \varphi)$). Finer discretization resolution of the microstructure domain $\Omega_0$ and axis space $S^2_{z+}$ leads to more accurate estimation of $P(D|\widehat{r})$, $\widehat{D}(\widehat{r})$, and $\widehat{H}(\widehat{r})$.

*3.1. Instantiation of discrete mesostructure*

A robust mesostructure modeling formalism based on grain shape idealization [11,14,82–84], which is implemented in the DREAM.3D code [85], was used to generate the mesostructure RVE. The RVE generation process assembles a sequence of modules (ellipsoidal grain generator, constrained grain packer, and seed point generator-constrained Voronoi tessellation) and produces a statistically realistic model of the corresponding polycrystalline aggregate based on the conventional morphological descriptors associated with grain shape idealization (Appendix A).

A cubic RVE with the edge length of 200 μm and $N_G = 1145$ grains was instantiated using the microstructure descriptors which were obtained from experimental microstructure characterization and analysis of an additively manufactured single-phase (austenitic) high-Mn steel as the model material [1,2]. The RVE was partitioned to a uniform grid of cubic elements of the same size (i.e., voxels) with the grid resolution of $N_v = 50^3$ voxels, where each voxel belongs to only one grain. The generated RVE features periodic boundaries: the grains that are cut by one of its faces appear from the opposite parallel face unless the respective face cuts the corresponding grain at its grain boundary. The instantiated RVE has the typical morphological features of AM metallic mesostructures including distinctive size heterogeneity and morphological texture corresponding to a pronounced columnar grain morphology.

It should be noted that the maximum number of grains embedded in a mesostructure volume element is limited by the extent of mesostructural complexities (i.e., morphological and crystallographic textures and heterogeneities) and the grid resolution (e.g., a grid resolution of $10^3$ cannot contain more than $10^3$ grains). In addition, the lower bound for the necessary number of grains in an RVE depends on the overall textures and heterogeneities. For example, to model a single-crystal material an RVE with a grid resolution of one voxel is sufficient, and to model a grain with an aspect ratio of 2 to 1, at least two voxels need to be allocated. The instantiated RVE was analyzed using the microstructure descriptors based on grain shape idealization to ensure that the morphological and crystallographic textures and heterogeneity characteristics of interest corresponding to typical AM metallic mesostructures were sufficiently captured.

Crystallographic orientations in terms of Bunge-Euler angles (Appendix D) were assigned to the grains of the RVE in such a way that no intragranular misorientation exists and the overall crystallographic orientation and misorientation distributions are almost random (i.e., absence of crystallographic texture). As a result, we assume that any anisotropy that arises in the macroscopic mechanical response must be due to the underlying anisotropic grain size of the grain ensemble. We adapted the iterative procedure described in Motaman et al. [1], which includes (i) random sampling of $N_G$ crystallographic orientations from the Bunge-Euler space, (ii) subsequent random assignment of the sampled orientations to the grains of the RVE, and (iii) repeating the random sampling and assignment steps until the norms of the corresponding crystallographic orientation distribution function (ODF) and correlated misorientation distribution function (MDF) (i.e., crystallographic orientation and misorientation texture indices) both fall below a prespecified threshold value. The crystallographic ODF and MDF in terms of Bunge-Euler angles (Appendix D) are computed using a fast algorithm for generalized spherical harmonics expansion based on kernel density estimation [86,87], which is implemented in the MATLAB®-based MTEX toolbox [88,89].

*3.2. Discretization of axis space*

Given a microstructure, to obtain the axial grain size distribution $P(D|\hat{r})$ and the corresponding grain size anisotropy $\widehat{D}(\hat{r})$ as well as the strain hardening anisotropy $\widehat{H}(\hat{r})$, one needs a uniform discretization of the axis space. A non-uniform discretization of the axis space would lead to biases: the regions on the surface of the unit sphere (i.e., the axis space), where the density of sampled points/axes is higher would have a higher weight in the corresponding distribution compared to the regions with lower density of sampled points. However, discretization of the surface of the unit sphere is not as straightforward as the uniform discretization of a cubic RVE embedded in the Euclidean 3-space. We adapted the algorithm proposed by Leopardi [90] to partition the surface of the northern hemisphere of the unit sphere ($S^2_{z+}$) into $N_r = 127 \in \mathbb{N}$ regions of equal area. The center point of each region, which is denoted by $\hat{r}_i$, represents the *i*-th axis used to measure the axial grain size and the uniaxial strain hardening. As such, $R \equiv \{\hat{r}_i \equiv (\theta_i, \varphi_i) \in S^2_{z+} \mid i \in \mathbb{N}; i \leq N_r\}$ represents the finite set of discrete measurement axes. Fig. 3 shows the projection plot of the grid of $N_r = 127$ equispaced axes $\hat{r}_i$ which were uniformly sampled from the surface of the northern hemisphere of the unit sphere ($S^2_{z+}$).

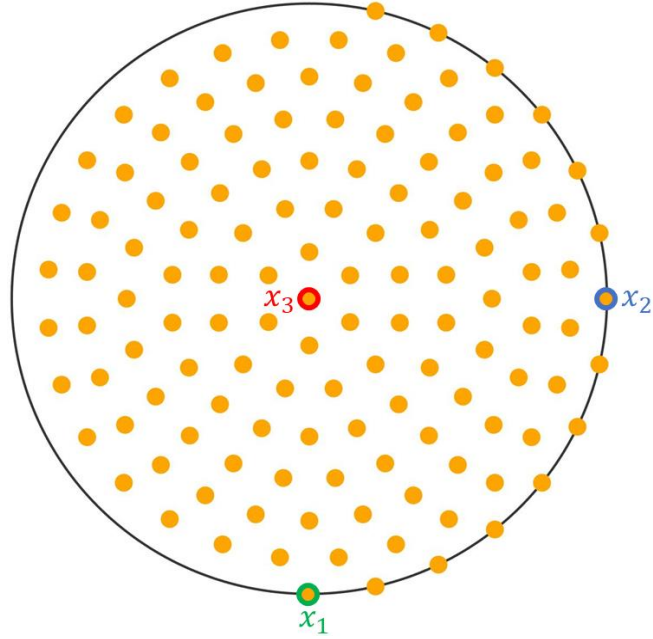

**Fig. 3.** Stereographic projection plot of the grid of 127 equispaced axes uniformly sampled from the surface of the northern hemisphere of the unit sphere ($S^2_{z+}$).

*3.3. Grain size anisotropy*

The instantiated RVE was used for axial grain size measurements as well as meso-scale full-field crystal plasticity simulations corresponding to uniaxial (tensile) deformation along the uniformly sampled axes. Prior to the axial grain size measurements, the instantiated cubic RVE was reconstructed so that the segments of the boundary grains (i.e., the grains cut by one of the faces of the RVE) with periodic counterparts were merged. The measurement of axial grain size was carried out at the center point of each voxel in the (undeformed) RVE. $X \equiv \{x_j \in \Omega_0 \mid j \in \mathbb{N}; j \leq N_v\}$ represents the set of measurement nodes associated with the discrete microstructure domain $\Omega_0$ and its corresponding $N_v$-voxel discretization, where $x_j$ denotes the position of the center of volume of the *j*-th voxel. Therefore, the measured axial grain size at the node $x_j \in X$ along the axis $\hat{r}_i \in R$, which is denoted by $D_{ij} \equiv D_{ij}(\hat{r}_i, x_j)$, can be obtained using Eqs. (5) or (9), where the indices $ij$ indicate that the value of the axial grain size is obtained by means of direct measurement. It should be noted that the grain boundaries in the instantiated discrete aggregate are comprised of voxel faces. Given Eqs. (12)-(15) and the sets $D_i \equiv D_i(\hat{r}_i) \equiv \{D_{ij} \mid j \in \mathbb{N}; j \leq N_v\}$ and $D \equiv \{D_{ij} \mid i,j \in \mathbb{N}; i \leq N_r; j \leq N_v\}$, the measured mean axial grain sizes, effective grain size, and the grain size anisotropy associated with the discrete microstructure domain $\Omega_0$ are calculated as follows:

$$\widehat{D}_i \equiv \widehat{D}_i(\hat{r}_i) \equiv \frac{\overline{D}_i}{\overline{\overline{D}}}; \quad \overline{D}_i \equiv \overline{D}_i(\hat{r}_i) \equiv \frac{1}{N_v}\sum_{j=1}^{N_v} D_{ij}; \quad \overline{\overline{D}} \equiv \frac{1}{N_v N_r}\sum_{i=1}^{N_r}\sum_{j=1}^{N_v} D_{ij} \equiv \frac{1}{N_r}\sum_{i=1}^{N_r}\overline{D}_i; \qquad (25)$$

where $\widehat{D}_i$ and $\overline{D}_i$, respectively, denote the measured grain size anisotropy and the measured mean axial grain size associated with the probe axis $\hat{r}_i$; and $\overline{\overline{D}}$ is the measured effective grain size.

Given Eq. (16), the measured discrete grain size anisotropies $\widehat{D}_i$ can be used to estimate the underlying continuous and smooth grain size anisotropy function $\widehat{D}(\hat{r})$ through finite series expansion of spherical harmonics as follows:

$$\widehat{D}(\hat{r}) \cong {}^K\widehat{D}(\hat{r}) = \sum_{l=0,2,4}^{K} \sum_{m=0}^{l} \left(\widetilde{D}_l^m Y_l^m(\hat{r}) + (-1)^{-m}\left(\widetilde{D}_l^m\right)^* Y_l^{-m}(\hat{r})\right); \quad (26)$$

where ${}^K\widehat{D}$ is the approximant of $\widehat{D}$ with the bandwidth or cut-off degree $K$. The procedure used for spherical harmonics approximation, which is estimating the spherical Fourier coefficients $\widetilde{D}_l^m$ using the measured discrete values $\widehat{D}_i$, is detailed in Appendix B. The bandwidth has a substantial impact on the accuracy of the approximant. By increasing the bandwidth, the approximant approaches to the measurement values: $\widehat{D}_i = \lim_{K \to \infty} {}^K\widehat{D}(\hat{r}_i)$. However, at large bandwidths, the sources of measurement noises due to the applied discretization penetrate to the approximant. Therefore, for a noise-reduced/denoised spherical harmonics approximation, the optimal bandwidth must be found. With an optimal bandwidth $K$, the approximant values ${}^K\widehat{D}(\hat{r}_i)$ are preferred to the measurement values $\widehat{D}_i \equiv \widehat{D}_i(\hat{r}_i)$ as they contain lower noise. In the following, we drop the superscript $K$ and use $\widehat{D}(\hat{r})$ and ${}^K\widehat{D}(\hat{r})$, interchangeably.

### 3.4. Crystal plasticity simulations and strain hardening anisotropy

The instantiated RVE (discrete microstructure domain $\Omega_0$) was used for meso-scale full-field crystal plasticity simulations corresponding to uniaxial tensile deformation along the sampled axes $\hat{r}_i \in R$. We used a physics-based crystal plasticity constitutive model (detailed in Appendix E) in the framework of the full-field method for computational polycrystal homogenization via a spectral solver based on fast Fourier transform, which was implemented in the modular crystal plasticity code DAMASK [1,91]. The adapted physics-based crystal plasticity constitutive model has been calibrated and validated for the model material (an additively manufactured austenitic high-Mn steel) in previous studies [1,2]. Note that the RVE was generated based on the results of microstructure analysis of the model material reported in the same studies. The parameters of the crystal plasticity constitutive model, which depend on temperature ($T = 298$ K), as well as the crystal symmetry, the processing history, and the mean chemical composition of the model material, are presented in Table E.1 (Appendix E).

The RVE was subjected to homogeneous macroscopic/far-field boundary conditions in terms of the rate of deformation gradient tensor $\dot{\overline{F}}$ and the first Piola-Kirchhoff stress tensor $\overline{P}$ (with complementary components) corresponding to uniaxial tensile deformation at constant temperature $T = 298$ K and constant macroscopic nominal strain rate $\dot{\varepsilon}_0 = 10^{-3}$ s$^{-1}$ along the sampled axes $\hat{r}_i \in R$ (see Appendix C and Section 3.2), which translate to pure displacement periodic boundary conditions on the RVE. The incrementally resolved fields at integration points are then homogenized over the mesoscopic RVE to render the macroscopic mechanical response in terms of the homogeneous macroscopic true strain and stress tensors $\overline{\varepsilon}$ and $\overline{\sigma}$ (Appendix C).

Each simulation was performed in the equivalent macroscopic true strain interval $[0, \varepsilon_f]$, where $\varepsilon_f = 0.35$. For each simulation, the equivalent macroscopic true stress-strain response ($\sigma(\varepsilon)$) was smoothed through Gaussian kernel density estimation and linearly interpolated at the fixed true strain increment $\Delta\varepsilon = 0.001$. This led to the discrete sequences of equivalent macroscopic true strain $\varepsilon_j = \varepsilon_{j-1} + \Delta\varepsilon$ (where $j \in \mathbb{N}$, $\varepsilon_0 = 0$, and $\varepsilon_j \in [0, \varepsilon_f])$ with the length of $N_\varepsilon \equiv \left\lfloor \frac{\varepsilon_f}{\Delta \varepsilon} \right\rfloor$ and the measured equivalent macroscopic true stress $\sigma_{ij} \equiv \sigma_{ij}(\hat{\boldsymbol{r}}_i, \varepsilon_j)$. Correspondingly, the measured instantaneous equivalent macroscopic strain hardening (or tangent modulus) was calculated by $H_{ij} \equiv H_{ij}(\hat{\boldsymbol{r}}_i, \varepsilon_j) \cong \frac{\sigma_{ij} - \sigma_{i(j-1)}}{\Delta \varepsilon}$. Analogous to Eq. (25) and based on Eqs. (19)-(21):

$$\widehat{H}_i \equiv \widehat{H}_i(\hat{\boldsymbol{r}}_i) \equiv \frac{\overline{H}_i}{\overline{\overline{H}}}; \quad \overline{H}_i \equiv \overline{H}_i(\hat{\boldsymbol{r}}_i) \equiv \frac{1}{N_\varepsilon} \sum_{j=1}^{N_\varepsilon} H_{ij}; \quad \overline{\overline{H}} \equiv \frac{1}{N_r N_\varepsilon} \sum_{i=1}^{N_r} \sum_{j=1}^{N_\varepsilon} H_{ij} \equiv \frac{1}{N_r} \sum_{i=1}^{N_r} \overline{H}_i; \tag{27}$$

where $\widehat{H}_i$ and $\overline{H}_i$, respectively, denote the measured strain hardening anisotropy and the measured mean uniaxial strain hardening associated with the probe axis $\hat{\boldsymbol{r}}_i$; and $\overline{\overline{H}}$ is the measured effective uniaxial strain hardening.

Given Eq. (22), the measured discrete uniaxial strain hardening anisotropies $\widehat{H}_i$ can be used to estimate the underlying continuous and smooth strain hardening anisotropy function $\widehat{H}(\hat{\boldsymbol{r}})$ through finite series expansion of spherical harmonic as follows:

$$\widehat{H}(\hat{\boldsymbol{r}}) \cong {}^K\widehat{H}(\hat{\boldsymbol{r}}) = \sum_{l=0,2,4}^{K} \sum_{m=0}^{l} \left( \widetilde{H}_l^m Y_l^m(\hat{\boldsymbol{r}}) + (-1)^{-m} \left(\widetilde{H}_l^m\right)^* Y_l^{-m}(\hat{\boldsymbol{r}}) \right); \tag{28}$$

where ${}^K\widehat{H}$ is the approximant of $\widehat{H}$ with the bandwidth $K$. The procedure used for spherical harmonics approximation, which is estimating the spherical Fourier coefficients $\widetilde{H}_l^m$ using the measured discrete values $\widehat{H}_i$, is detailed in Appendix B. By increasing the bandwidth, the approximant ${}^K\widehat{H}(\hat{\boldsymbol{r}}_i)$ approaches to the measurement values $\widehat{H}_i \equiv \widehat{H}_i(\hat{\boldsymbol{r}}_i)$. Nevertheless, at large bandwidths, the measurement noises will be included in the approximant. With an optimal bandwidth $K$, the (denoised) approximant values ${}^K\widehat{H}(\hat{\boldsymbol{r}}_i)$ are preferred to the measurement values $\widehat{H}_i \equiv \widehat{H}_i(\hat{\boldsymbol{r}}_i)$ as they contain lower noise. Henceforth, we drop the superscript $K$ and use $\widehat{H}(\hat{\boldsymbol{r}})$ and ${}^K\widehat{H}(\hat{\boldsymbol{r}})$, interchangeably.

## 4. Results and discussion

### 4.1. Instantiated RVE

As mentioned in Section 3.1, a discrete mesostructure was generated as a cubic RVE with periodic boundaries, edge length of 200 μm, 1145 grains, and grid resolution of $50^3$ voxels. The RVE was instantiated using the conventional microstructure descriptors associated with grain shape idealization (Appendix A), which were derived via experimental microstructure characterization and analysis of the model material (an AM austenitic high-Mn steel) [1,2]. The instantiated RVE has grain size heterogeneity and morphological texture corresponding to a pronounced columnar grain morphology typically observed in AM metals. Crystallographic orientations were assigned to the grains of the RVE in a such way to obtain almost random crystallographic orientation and misorientation distributions.

Fig. 4 shows the grain structure and inverse pole figure (IPF) orientation map of the generated RVE. As shown in Fig. 4, most of the grains in the RVE are elongated along the build direction (BD) and the scan direction (SD) relative to the transverse direction (TD). The distributions of the conventional morphological descriptors associated with the generated RVE are shown in Fig. 5. The means of the distributions of the conventional morphological descriptors corresponding to the instantiated RVE are presented in Table 2. As shown in Fig. 5a, the instantiated RVE has a heterogeneous multi-modal equivalent grain size distribution. The notable deviation between the number- and volume-weighted mean equivalent grain sizes (Table 2) is an indicator of the highly heterogeneous grain size distribution present in the RVE. The extremely low mean equivalent aspect ratios imply the prevalence of highly elongated grain shapes. The equivalent shape-axes orientation distribution reveals that the grains in the RVE are dominantly oriented in such a way that their equivalent major semi-axes ($\hat{e}_a$) and equivalent intermediate semi-axes ($\hat{e}_b$) have low angles relative to $x_3$ (BD) and $x_1$ (SD), respectively. The norm of the equivalent axes ODF (i.e., morphological orientation texture index) associated with the RVE is 36.3.

These observations indicate that the generated RVE features a pronounced columnar grain morphology resembling the typical microstructures of AM metallic materials. However, it is important to exercise caution when interpreting the distributions of the conventional morphological descriptors. These distributions do not conclusively confirm the presence of collective grain size anisotropy (i.e., collective anisotropy of grain boundary spacing) in the microstructure. Furthermore, by solely considering the distributions of the conventional morphological descriptors without any information about their correlations, as depicted in Fig. 5, one cannot accurately assess the extent of collective grain size anisotropy in the microstructure.

Fig. 6 shows the crystallographic orientation distribution (in terms of IPFs) and the crystallographic misorientation distribution (in axis-angle convention) associated with the instantiated RVE. As shown in Fig. 6a, the RVE has a nearly random crystallographic orientation distribution. In Fig. 6b and c, the (correlated) crystallographic misorientation angle and axis distributions in the RVE are compared to their associated uncorrelated/Mackenzie misorientation angle and axis distributions. The negligible difference between the correlated and uncorrelated misorientation distributions implies an almost random crystallographic misorientation distribution in the RVE. The crystallographic orientation and misorientation texture indices (norms of crystallographic ODF and correlated MDF) associated with the RVE are 1.039 and 1.007, respectively. Hence, while the RVE is morphologically strongly textured, it is crystallographically almost textureless. As a result, any notable anisotropy in the macroscopic mechanical response of the RVE is due to its grain morphology.

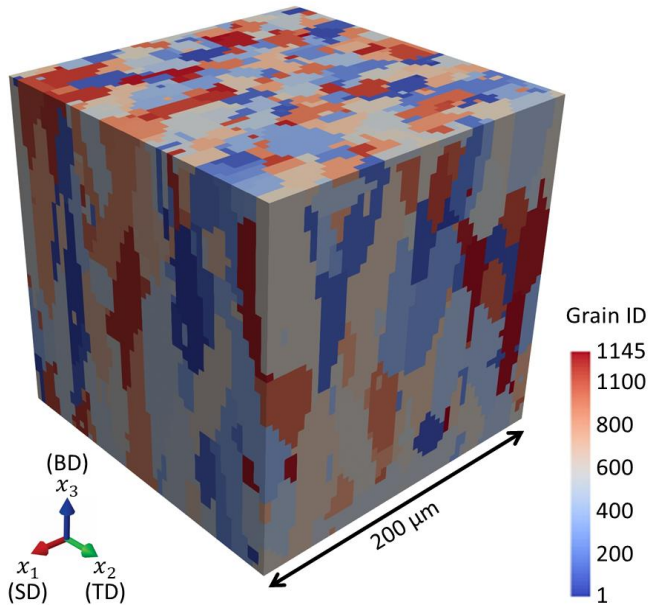
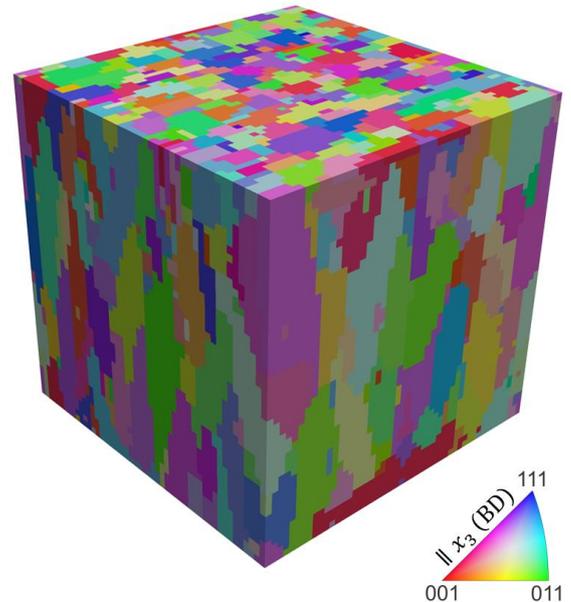

**Fig. 4.** The instantiated cubic RVE with periodic boundaries, edge length of 200 μm, 1145 grains, and grid resolution of $50^3$ voxels, which has the typical grain size heterogeneity and columnar grain morphology associated with AM metals.

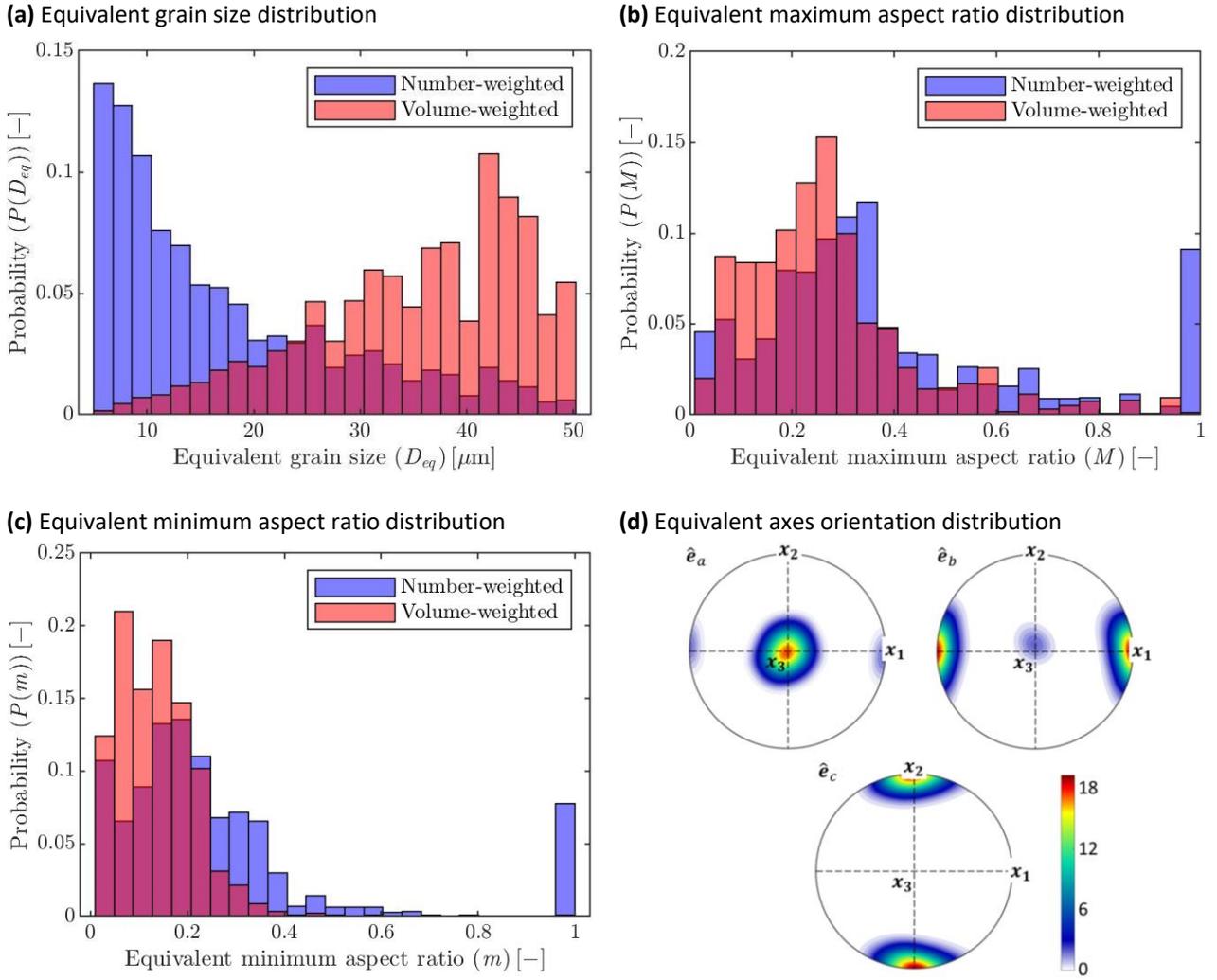

**Fig. 5.** Distributions of the conventional morphological descriptors (Appendix A) associated with the generated RVE. The equivalent axes orientation distribution (d) is represented in terms of pole figures with respect to grains equivalent (ellipsoidal) semi-axes $\{\hat{\boldsymbol{e}}_a, \hat{\boldsymbol{e}}_b, \hat{\boldsymbol{e}}_c\}$.

**Table 2.** The means of the conventional morphological descriptors associated with the generated RVE.

| Conventional morphological descriptor | Number-weighted mean | Volume-weighted mean |
| --- | --- | --- |
| Equivalent grain size | 17.4 μm | 35.3 μm |
| Equivalent minimum aspect ratio | 0.262 | 0.137 |
| Equivalent maximum aspect ratio | 0.378 | 0.271 |

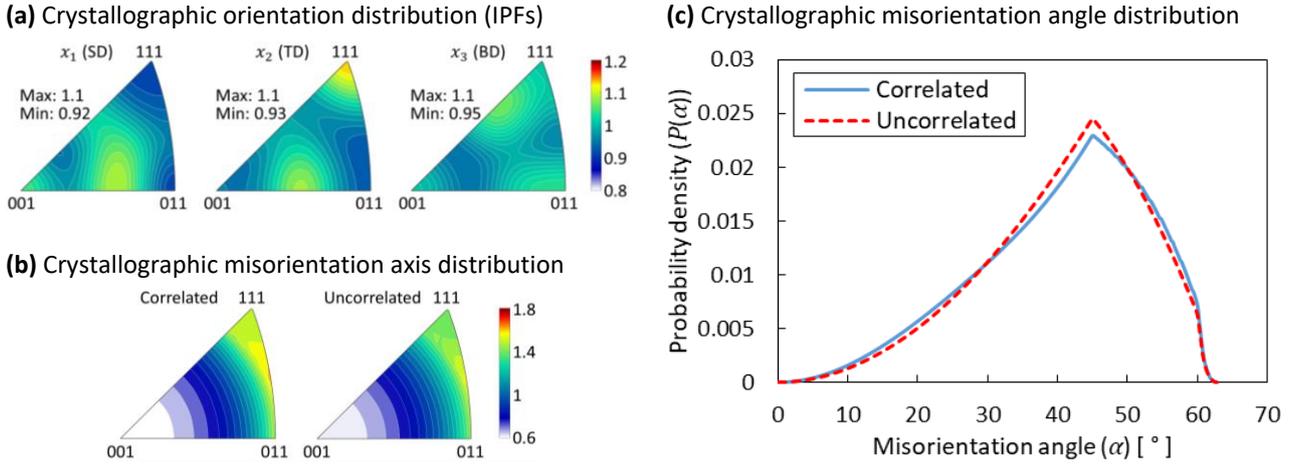

**Fig. 6.** Crystallographic orientation (volume-weighted) distribution (in terms of IPFs) and crystallographic misorientation distribution (in axis-angle convention) associated with the instantiated RVE.

*4.2. Grain size anisotropy*

The axial grain sizes $D_{ij} \equiv D_{ij}(\hat{r}_i, x_j)$ and the mean axial grain sizes $\bar{D}_i \equiv \bar{D}_i(\hat{r}_i)$ were measured along the uniformly sampled axes $\hat{r}_i \in R$ at the center point of each voxel $x_j \in X$ in the RVE using Eqs. (5) or (9). The distribution of the measured axial grain sizes $D_{ij}$ (i.e., grain boundary spacings) in the instantiated RVE is shown in Fig. 7. The mean of the distribution of the measured axial grain sizes yields the measured effective grain size $\bar{\bar{D}} = 19.1$ μm. The measured effective grain size was used to calculate the measured grain size anisotropies $\hat{D}_i$ corresponding to the sampled axes $\hat{r}_i \in R$ according to Eq. (25). Subsequently, based on the procedure detailed in Appendix B and according to Eq. (26), the underlying grain size anisotropy function $\hat{D}(\hat{r})$ associated with the generated RVE was approximated by a finite series expansion of spherical harmonics with the bandwidth $K = 10$. Fig. 8 shows the stereographic projection plots of the measured grain size anisotropies $\hat{D}_i$ and the corresponding approximated grain size anisotropy function $\hat{D}(\hat{r})$ associated with the generated RVE.

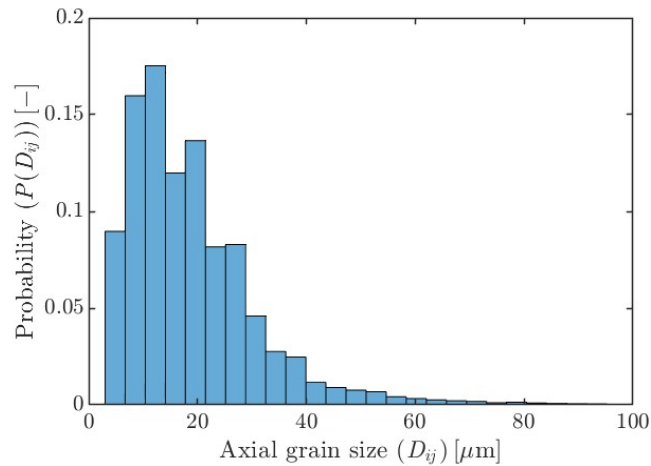

**Fig. 7.** Distribution histogram of $15.875 \times 10^6$ measured axial grain sizes in the instantiated RVE. The mean of the distribution corresponds to the measured effective grain size $\bar{\bar{D}} = 19.1$ μm.

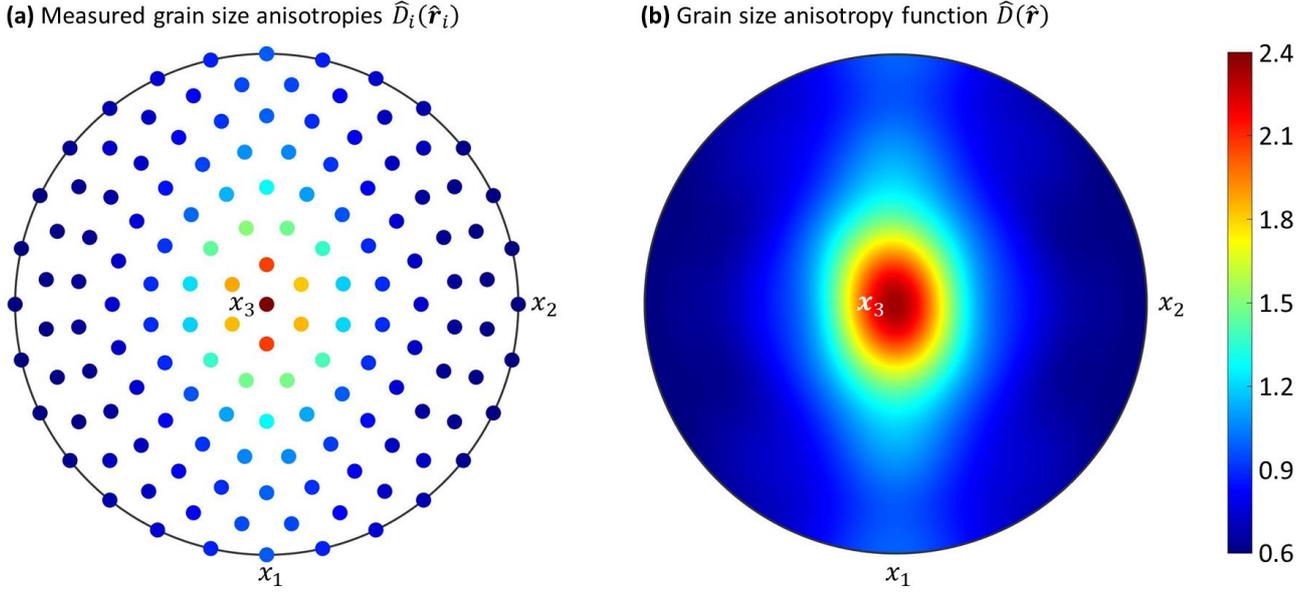

**Fig. 8.** Stereographic projection plots of (a) the measured grain size anisotropy at the sampled axes $\hat{\boldsymbol{r}}_i \in R$, and (b) the corresponding approximated grain size anisotropy function associated with the generated RVE.

The measurement noises corresponding to the high-frequency spherical harmonics of degree $l > K$ with relatively low amplitude are filtered in the (approximated) grain size anisotropy function. The root-mean-square deviation (RMSD) between the grain size anisotropy function ($\widehat{D}(\hat{\boldsymbol{r}}_i)$) and the measured grain size anisotropy corresponding to the axes $\hat{\boldsymbol{r}}_i \in R$ ($\widehat{D}_i$) is relatively low: $\sqrt{\frac{1}{N_r}\sum_{i=1}^{N_r}\left(\widehat{D}(\hat{\boldsymbol{r}}_i) - \widehat{D}_i(\hat{\boldsymbol{r}}_i)\right)^2} = 0.021$. The measurement noise/error sources are associated with the discretization of the microstructure domain and can be divided into two categories: (i) those related to the resolution of the discretization (in other words, coarser resolution results in higher noise), and (ii) those stemming from the shape anisotropy of the discretization elements (e.g., cubic voxels exhibit a certain shape anisotropy [2]).

The grain size anisotropy function ($\widehat{D}$) effectively represents the collective anisotropy of grain boundary spacing of microstructure. The stereographic projection plot of the continuous grain size anisotropy function shown in Fig. 8b is the signature of columnar grain morphology. The generated RVE features a strongly anisotropic grain size, which is suggested by its grain size anisotropy index $\mathcal{A}[\widehat{D}] = 1.131$. As shown in Fig. 8, the grain size anisotropy of the axis $x_3$ (BD) has the highest value ($\widehat{D}(\hat{\boldsymbol{e}}_3) \approx 2.4$), while the axes oriented closely to $x_2$ (TD) have relatively low grain size anisotropy ($\widehat{D}(\hat{\boldsymbol{e}}_2) \approx 0.6$). This indicates that the mean axial grain size corresponding to the axis $x_3$ is about four times higher than that associated with the axis $x_2$. Moreover, the grain size anisotropy of the axis $x_1$ (SD) ($\widehat{D}(\hat{\boldsymbol{e}}_1) \approx 1$) is between those of the other principal axes of the reference frame: $\widehat{D}(\hat{\boldsymbol{e}}_3) > \widehat{D}(\hat{\boldsymbol{e}}_1) > \widehat{D}(\hat{\boldsymbol{e}}_2)$. This is consistent with the conventional equivalent axes orientation distribution shown in Fig. 5d.

Furthermore, as mentioned in Section 2.4, each continuous anisotropy function $\widehat{D}(\hat{\boldsymbol{r}})$ uniquely defines a 3D shape. Therefore, the instantiated RVE has an equivalent grain shape represented by its grain size anisotropy function $\widehat{D}(\hat{\boldsymbol{r}})$. Fig. 9 shows the principal orthogonal sections of the equivalent grain shape of the generated RVE. As shown in Fig. 9, the RVE has a near ellipsoidal equivalent grain shape whose major, intermediate, and minor semi-axes are aligned with the principal axes $x_3$, $x_1$, and $x_2$, respectively, and its maximum and minimum aspect ratios are roughly 0.41 and 0.25. This was somewhat expected as the first step in the algorithm used for RVE instantiation is the generation of ellipsoidal grains (Section 3.1).

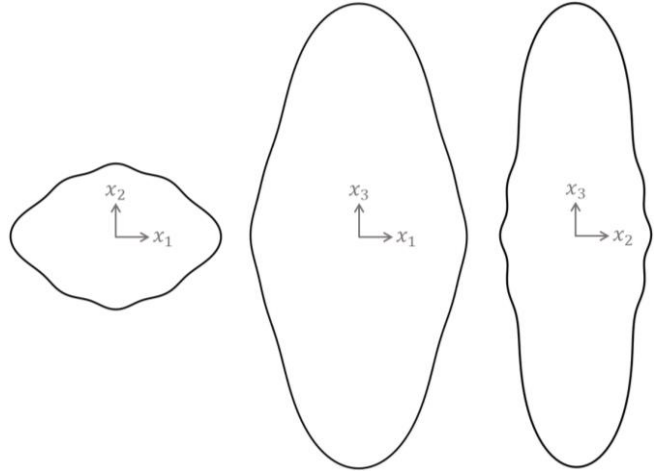

**Fig. 9.** Principal orthogonal sections of the equivalent grain shape of the generated RVE, which are derived from its grain size anisotropy function.

### 4.3. Strain hardening anisotropy

Fig. 10 shows the equivalent macroscopic true stress and strain hardening responses of the RVE under homogeneous macroscopic/far-field boundary conditions corresponding to uniaxial tension at constant temperature $T = 298$ K and constant macroscopic nominal strain rate $\dot{\varepsilon}_0 = 10^{-3}$ s$^{-1}$ along the uniformly sampled axes $\hat{\boldsymbol{r}}_i \in \mathrm{R}$. In the following, for brevity, we drop the term "equivalent macroscopic" as the common adjective of true strain, true stress, and strain hardening.

**(a)** Equivalent macroscopic true stress response        **(b)** Equivalent macroscopic strain hardening response

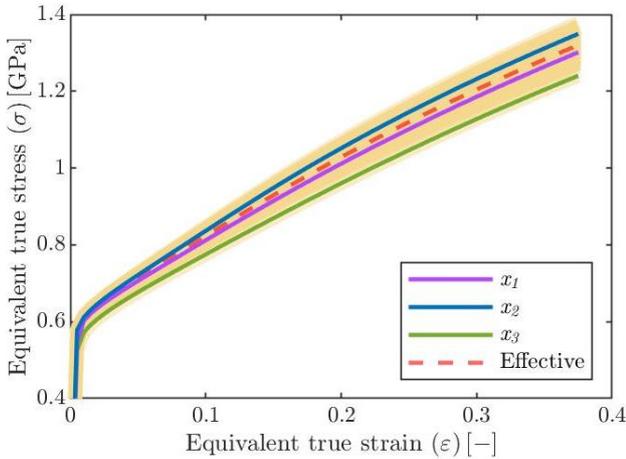 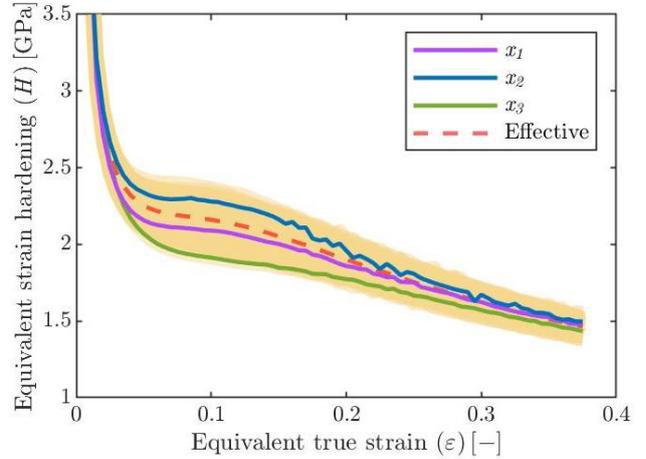

**Fig. 10.** Equivalent macroscopic true stress and strain hardening responses of the instantiated RVE under homogeneous macroscopic/far-field boundary conditions corresponding to uniaxial tension at constant temperature $T = 298$ K and constant macroscopic nominal strain rate $\dot{\varepsilon}_0 = 10^{-3}$ s$^{-1}$ along the uniformly sampled axes $\hat{\boldsymbol{r}}_i \in \mathrm{R}$. In each figure, 127 orange-colored curves representing the macroscopic mechanical responses associated with different axes $\hat{\boldsymbol{r}}_i \in \mathrm{R}$ are plotted.

As shown in Fig. 10, there is a significant variation in the macroscopic mechanical response of the RVE associated with different tensile axes, indicating a highly anisotropic macroscopic mechanical response. While the spread of true stress at small true strains is relatively low, it is significant at large strains. At true strain $\varepsilon = \varepsilon_f = 0.35$, which corresponds to the onset of necking in the model material, the spread of true stress is approximately 200 MPa. In contrast, the spread of uniaxial strain hardening responses at small strains is higher compared to large strains. At true strain $\varepsilon = 0.1$, the spread in uniaxial strain hardening responses exceeds 600

MPa. In fact, the relatively high initial spread in uniaxial strain hardening responses is the reason for the divergence of true stress as true strain increases. As shown in Fig. 10, the lower bound of the true stress and strain hardening responses for the sampled axes $\hat{r}_i \in R$ corresponds to the tensile load axis $x_3$, while the axis $x_2$ is nearly the upper bound. The effective true stress and strain hardening responses plotted in Fig. 10 represent the average of those associated with all the sampled axes $\hat{r}_i \in R$. Hence, one may consider the effective response as the hypothetical isotropic uniaxial mechanical response of the corresponding polycrystal in the complete absence of crystallographic and morphological textures.

Given the uniaxial strain hardening responses of the RVE and according to Eq. (27), the measured effective uniaxial strain hardening ($\bar{\bar{H}} \approx 1.91$ GPa) as well as the measured mean uniaxial strain hardenings ($\bar{H}_i \equiv \bar{H}_i(\hat{r}_i)$) and the measured strain hardening anisotropies ($\hat{H}_i \equiv \hat{H}_i(\hat{r}_i)$) associated with the sampled axes $\hat{r}_i \in R$ are calculated. Then, based on the procedure described in Appendix B and Eq. (28), the underlying strain hardening anisotropy function $\hat{H}(\hat{r})$ associated with the generated RVE was approximated by a finite series expansion of spherical harmonics with the bandwidth $K = 4$. Fig. 11 shows stereographic projection plots of the measured strain hardening anisotropies $\hat{H}_i$ and the corresponding approximated strain hardening anisotropy function $\hat{H}(\hat{r})$ associated with the generated RVE.

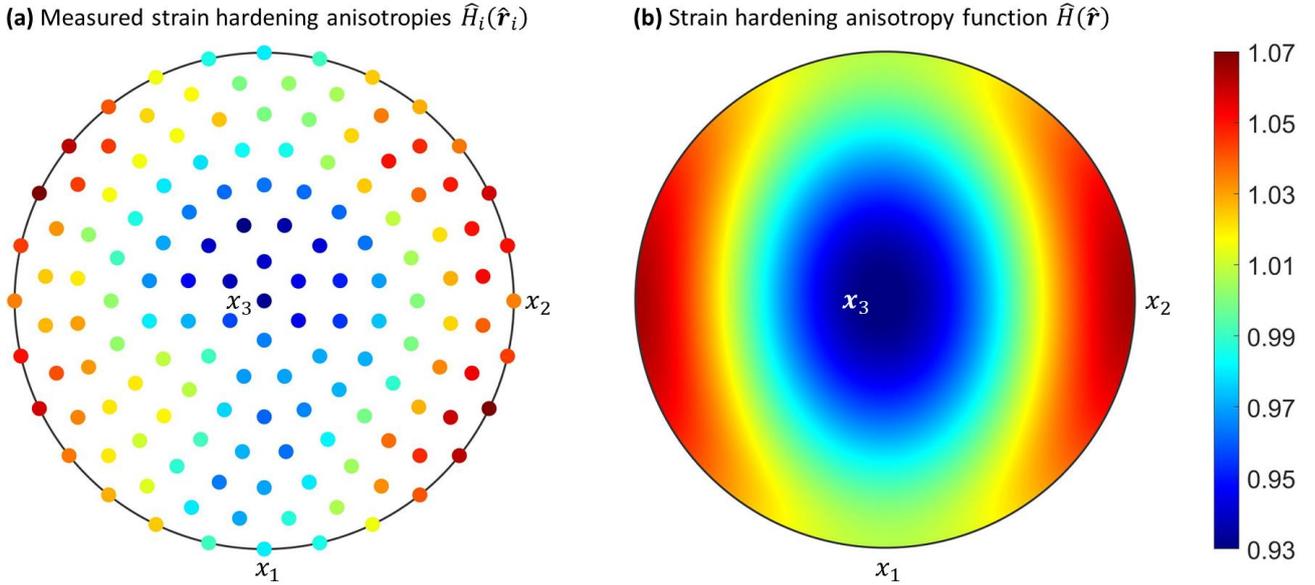

**Fig. 11.** Stereographic projection plots of (a) the measured strain hardening anisotropy at the sampled axes $\hat{r}_i \in R$, and (b) the corresponding approximated strain hardening anisotropy function associated with the generated RVE.

To filter the notable noise in the measured strain hardening anisotropy data (Fig. 11a), a relatively low bandwidth ($K = 4$) was chosen. Accordingly, the measurement noises corresponding to the high-frequency spherical harmonics of degree $l > K$ with low amplitude are filtered in the (approximated) strain hardening anisotropy function. The root-mean-square deviation (RMSD) between the strain hardening anisotropy function ($\hat{H}(\hat{r}_i)$) and the measured strain hardening anisotropy corresponding to the axes $\hat{r}_i \in R$ ($\hat{H}_i(\hat{r}_i)$) is $\sqrt{\frac{1}{N_r}\sum_{i=1}^{N_r}\left(\hat{H}(\hat{r}_i) - \hat{H}_i(\hat{r}_i)\right)^2} = 0.015$. The noise sources in the grain size anisotropy measurements (Section 4.2) also add noise to the strain hardening anisotropy measurements. However, there are two additional sources of noise in the strain hardening anisotropy measurements: (i) errors in the crystal plasticity constitutive modeling and simulation, and most importantly, (ii) noise in the crystallographic orientation and misorientation distributions in the RVE. Owing to the presence of these extra noise sources, the measurement noise in the measured strain hardening anisotropies $\hat{H}_i$ is considerably higher than that in the measured grain size anisotropies $\hat{D}_i$. Consequently, to mitigate the increased noise in the measured strain hardening anisotropies, the optimal bandwidth for the spherical harmonics approximation of the strain hardening anisotropy function $\widehat{H}(\widehat{r})$ is lower than that of the grain size anisotropy function $\widehat{D}(\widehat{r})$.

As depicted in Fig. 6, the crystallographic orientation and misorientation distributions in the RVE are nearly but not perfectly random as there exist local fluctuations around the corresponding fully random distributions. These local fluctuations in the crystallographic orientation and misorientation distributions and the resulting deviations from the corresponding perfectly random distributions give rise to anisotropy fluctuations in the macroscopic mechanical response of the RVE. In other words, a portion of the measured strain hardening anisotropy is attributed to the deviation of crystallographic orientation and misorientation distributions from complete randomness. Since achieving a completely textureless RVE in terms of crystallography is unfeasible, denoising the measured strain hardening anisotropies plays a crucial role in isolating the effect of morphological texture on the macroscopic mechanical response of the RVE from that associated with crystallographic texture. For instance, the lower measured strain hardening anisotropy for the axis $x_2$ in comparison to its nearby axes (Fig. 11a) can be explained by the relatively higher intensity of <111> ∥ $x_2$ in the crystallographic orientation distribution within the RVE (Fig. 6a). As demonstrated in Fig. 11b, the axis $x_2$ represents the exact lower bound of the denoised strain hardening anisotropy function.

The uniaxial strain hardening response of the generated RVE is anisotropic, which is suggested by its strain hardening anisotropy index $\mathcal{A}[\widehat{H}] = 0.102$. As shown in Fig. 11, the strain hardening anisotropy associated with the axis $x_3$ (BD) has the lowest value ($\widehat{H}(\widehat{e}_3) \approx 0.93$), while the axes oriented closely to $x_2$ (TD) have relatively high strain hardening anisotropy ($\widehat{H}(\widehat{e}_2) \approx 1.07$). Even though these values are lower compared to those associated with the grain size anisotropy in the RVE, they represent a considerable spread in the uniaxial mechanical response of the RVE as shown in Fig. 10. Moreover, the strain hardening anisotropy corresponding to the axis $x_1$ (SD) ($\widehat{H}(\widehat{e}_1) \approx 1$) is between those of the other principal axes of the reference frame: $\widehat{H}(\widehat{e}_3) < \widehat{H}(\widehat{e}_1) < \widehat{H}(\widehat{e}_2)$.

*4.4. Anisotropic strain hardening due to anisotropic grain size*

Since the crystallographic orientation and misorientation distributions in the instantiated RVE are almost random (i.e., crystallographically textureless), the main source of the observed strain hardening anisotropy is the grain size anisotropy (i.e., morphological texture). By comparing Fig. 8 and Fig. 11, we find that the axes exhibiting relatively high grain size anisotropy correspond to relatively low strain hardening anisotropy, and vice versa. In other words, a relatively large mean axial grain size along an axis results in a relatively low mean uniaxial strain hardening along that axis; and, conversely, a relatively small mean axial grain size along a certain axis leads to a relatively high mean uniaxial strain hardening along that axis. These trends can be clearly seen in the measurement dataset $\left\{\left(\widehat{D}_i(\widehat{r}_i), \widehat{H}_i(\widehat{r}_i)\right) \middle| \widehat{r}_i \in \mathrm{R}\right\}$ and the denoised dataset $\left\{\left(\widehat{D}(\widehat{r}_i), \widehat{H}(\widehat{r}_i)\right) \middle| \widehat{r}_i \in \mathrm{R}\right\}$ plotted in Fig. 12.

As shown in Fig. 12, the following inverse square model, which describes the strain hardening anisotropy due to grain size anisotropy, can be fitted to both the measurement and denoised datasets:

$$\widehat{H}(\widehat{r}) = k \left(\widehat{D}(\widehat{r})\right)^{-2} + h; \tag{29}$$

where $k$ and $h$ are two dimensionless constants. As shown in Fig. 12, the coefficient of determination ($R^2$) is improved from 0.81 to 0.97 after denoising the measurement dataset. The model predicts monotonically decreasing strain hardening anisotropy by increasing grain size anisotropy according to an inverse square relation. In the absence of crystallographic texture, if the mean axial grain size is isotropic ($\forall \widehat{r} \in S^2: \widehat{D}(\widehat{r}) =$ 1), one would expect isotropic mean uniaxial strain hardening ($\forall \hat{r} \in S^2: \hat{H}(\hat{r}) = 1$). Inserting $\hat{D}(\hat{r}) = 1$ and $\hat{H}(\hat{r}) = 1$ into Eq. (29) leads to $k + h = 1$, implying that, theoretically, the constants $k$ and $h$ are not independent. The best fit of the model to the denoised dataset (Fig. 12b) corresponds to $k = 0.06$ and $h = 0.93$ ($k + h = 0.99 \approx 1$). This deviation between the theoretical $h + k = 1$, and the calibrated $h + k = 0.99$ is associated with the errors accumulated in the measurements of the effective grain size $\bar{\bar{D}}$ and the effective strain hardening $\bar{\bar{H}}$, which serve as normalization factors for the grain size anisotropy and strain hardening anisotropy functions, respectively. Through a recalibration and renormalization procedure, one can correct and minimize the error in the measurements of the effective grain size and the effective strain hardening and satisfy the strict requirement that $h + k = 1$.

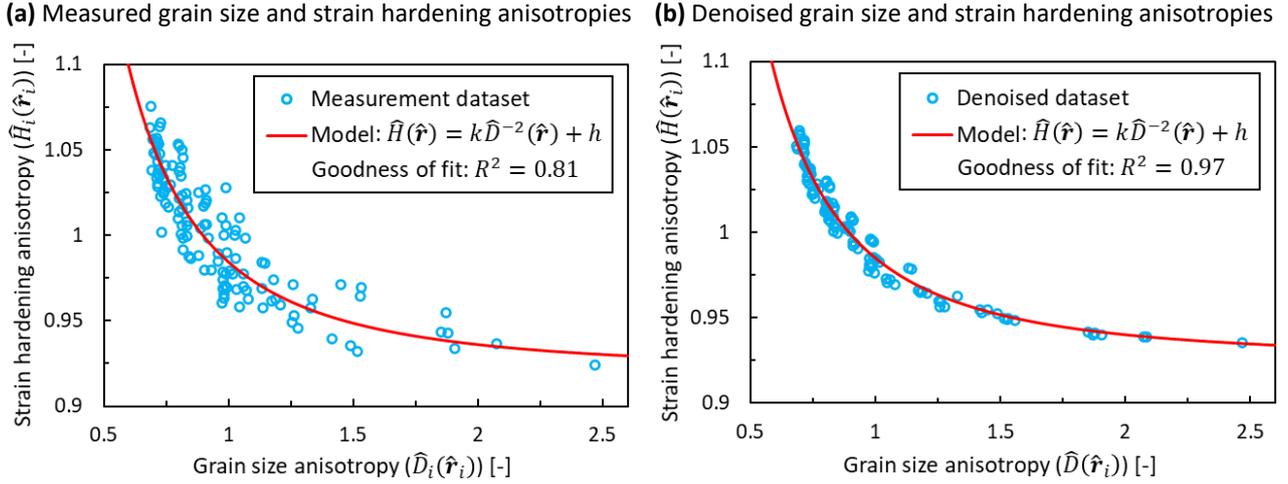

**Fig. 12.** Measured and denoised strain hardening anisotropies vs. grain size anisotropies. The superposed inverse square model was fitted to each dataset. The best fit of the model to the denoised dataset corresponds to $k = 0.06$ and $h = 0.93$.

The axial grain size distribution (Fig. 7) as well as the number- and volume-weighted distributions of equivalent grain size (Fig. 5a and Table 2) of the modeled RVE (Fig. 4) all indicate an effective grain size greater than 15 µm. Analysis of electron backscatter diffraction (EBSD) orientation maps of two orthogonal sections of the model material (AM single-phase austenitic high-Mn steel) also suggest an effective grain size greater than 15 µm [1,2]. According to the constitutive parameters incorporated into the crystal plasticity constitutive model (Table E.1 in Appendix E), the initial (total) dislocation density of the model material is $\rho_{t0} = 3.72 \times 10^{14}$ m$^{-2}$. Analysis of EBSD orientation maps and TEM micrographs of the model material indicate a relatively high initial dislocation density in the order of $10^{14}$ m$^{-2}$ [1,2]. In addition, the experimental testing of the model material in the aforementioned previous studies as well as the macroscopic mechanical responses obtained by crystal plasticity simulations in the present study (Fig. 10) all show a continuous yielding behavior.

Similar to the model material in the present study, as-built AM metallic materials often feature relatively high dislocation density and/or coarse grains, and typically exhibit a continuous yielding behavior. Consequently, the grain size dependence of the macroscopic mechanical response of the adapted model material and most of the (as-built) AM metals is governed by the grain boundary hardening mechanisms leading to enhanced strain hardening near grain boundaries (Appendix F). In this regime of the grain size dependence of the macroscopic mechanical response of polycrystals, characterized by continuous yielding response and dominance of grain boundary hardening mechanisms, the equivalent form the Hall-Petch relation, which describes the relationship between the effective grain size and the effective strain hardening, is preferable (Eq. (F.2) in Appendix F). This relation conveys that a smaller effective grain size results in higher (macroscopic) effective strain hardening. Analogously, Eq. (29) suggests that, in a crystallographically textureless single-phase polycrystal, a relatively small mean axial grain size along a load axis leads to a relatively high mean uniaxial strain hardening along that axis, and vice versa.

Even though Eq. (29) and the Hall-Petch relation (Eqs. (F.1) and (F.2) in Appendix F) share obvious similarities, it should be noted that the Hall-Petch scaling exponents and the scaling exponent in Eq. (29) are not necessarily the same (Appendix F). The proposed relatively low values for the Hall-Petch scaling exponent compared to $\widehat{D}^{-2}$ term in Eq. (29) can be explained by the fact that the initial yield stress ($\bar{\bar{\sigma}}_y$) or the effective strain hardening ($\bar{\bar{H}}$) in the Hall-Petch relation correspond to small strains, whereas $\widehat{H}(\widehat{r})$ in Eq. (29) is associated with the large strain $\varepsilon = \varepsilon_f$ (Section 3.4). Furthermore, despite the similarities, the Hall-Petch relation and Eq. (29) are independent relations that describe different aspects of the microstructure-properties relationship in polycrystals. As mentioned, the classical Hall-Petch relation is isotropic. Nevertheless, regardless of the presence of grain size and strain hardening anisotropies, the generalized Hall-Petch relation (Eqs. (F.1) and (F.2) in Appendix F) links the initial yield stress ($\bar{\bar{\sigma}}_y$) or the effective strain hardening ($\bar{\bar{H}}$) at small strains to the effective grain size $\bar{\bar{D}}$. In contrast, Eq. (29) describes the contribution of grain size anisotropy to the strain hardening anisotropy and thus is a relationship between two anisotropy functions. By inserting Eqs. (15) and (21) in Eq. (29), we obtain the following equation which relates the mean uniaxial strain hardening $\bar{H}(\widehat{r})$ to the mean axial grain size $\bar{D}(\widehat{r})$ associated with the axis $\widehat{r}$:

$$\bar{H}(\widehat{r}) = k\bar{\bar{H}}\bar{\bar{D}}^2\big(\bar{D}(\widehat{r})\big)^{-2} + h\bar{\bar{H}}; \tag{30}$$

where the effective strain hardening $\bar{\bar{H}}$ can be replaced by the Hall-Petch relation (Eq. (F.2) in Appendix F), so that the mean uniaxial strain hardening $\bar{H}(\widehat{r})$ associated with the axis $\widehat{r}$ can be expressed only in terms of the mean axial grain size $\bar{D}(\widehat{r})$ and the effective grain size $\bar{\bar{D}}$: $\bar{H}(\widehat{r}) = \big(k_0\bar{\bar{D}}^n + h_0\big)\big(k\bar{\bar{D}}^2\big(\bar{D}(\widehat{r})\big)^{-2} + h\big)$. Hence, Eq. (29) and the Hall-Petch equation are two complementary microstructure-properties relationships. It is conceivable that a crystallographically textureless single-phase polycrystal having the same mean axial grain size along each and every axis (i.e., isotropic mean axial grain size) exhibits the same mean uniaxial strain hardening along any load axis (i.e., isotropic response). In a crystallographically textureless polycrystal with isotropic mean axial grain size ($\forall \widehat{r} \in S^2: \bar{D}(\widehat{r}) = \bar{\bar{D}}, \bar{H}(\widehat{r}) = \bar{\bar{H}}$), both sides of Eq. (30) will vanish after trivial rearrangements (i.e., $0 = 0$). Therefore, in an isotropic polycrystal, unlike the Hall-Petch relation, Eqs. (29) and (30) do not provide any useful information.

The grain size dependence of the macroscopic mechanical response of the adapted model material, similar to most AM metals, falls under the first regime of the grain size effect (Appendix F), which is characterized by the predominance of grain boundary hardening mechanisms including grain boundary dislocation pileups and sources as well as elasto-plastic deformation incompatibilities at grain boundaries. As the mean axial grain size corresponding to an axis decreases, the grain boundary surface area density normal to that axis increases. This leads to a higher volumetric frequency of the grain boundary hardening mechanisms during uniaxial deformation along that (load) axis and hence a higher mean uniaxial strain hardening corresponding to the axis (according to the homogenization law in Eq. (C.6)). This observation is consistent with Eq. (29). The aforementioned grain boundary hardening mechanisms are incorporated into the adapted full-field crystal plasticity model. Grain boundaries, the interaction between contiguous grains, and thus elasto-plastic deformation incompatibilities at grain boundaries are explicitly taken into account through the RVE and the adapted full-field method for computational polycrystal homogenization.

In addition, the enhanced strain hardening near grain boundaries due to grain boundary dislocation sources and pileups are integrated into the physics-based crystal plasticity constitutive model (Appendix E). In particular, the term $\frac{1}{D_m(\widehat{\underline{b}}_{sl}^\alpha, x)}$ in the slip mean free path relation (Eq. (E.16) in Appendix E) guarantees that the material points in the vicinity of grain boundaries experience a proliferation of dislocation multiplication at the onset of plastic deformation through increasing the number density of grain boundary sources (at boundary dislocations, and disconnections such as ledges, steps, kinks, and trijunctions). This entails a higher dislocation accumulation rate (Eq. (E.13) in Appendix E) and hence formation of dislocation pileups near grain boundaries, which leads to a shorter length of mobile dislocation segments (due to shorter effective forest spacing) (Eq. (E.17) in Appendix E) and a higher strain hardening in the grain boundary regions throughout the plastic deformation.

The significance of Eq. (29) lies in its potential to incorporate macroscopic anisotropy (originating from microstructural textures) into physics-based macro-scale constitutive models of polycrystal plasticity [92,93] for efficient simulation of structure-performance linkage, which plays an essential role in optimal design for metal additive manufacturing [94]. It is presumed that any single-phase polycrystal, which lacks both morphological texture and crystallographic texture, exhibits an isotropic mechanical response irrespective of other microstructural aspects. A previous study on the adapted model material concluded that both morphological texture (i.e., grain size anisotropy) and crystallographic texture almost equally affect the anisotropic strain hardening behavior, and the overall macroscopic strain hardening anisotropy is the superposition of the individual contributions of morphological and crystallographic textures [1]. Nevertheless, the primary focus of the present study is to investigate the impact of morphological texture on strain hardening anisotropy. This elucidates why we assigned crystallographic orientations to the grain of the instantiated RVE in a way to obtain random distributions of crystallographic orientation and misorientation (Section 3.1).

## 5. Outlook

This study is based on a systematic numerical analysis comprising the instantiation of an RVE of a polycrystalline aggregate characterized by strongly anisotropic grain size, the measurement of axial grain sizes within the RVE along various axes, and full-field crystal plasticity simulations corresponding to uniaxial tensile deformation along those same axes. The RVE was generated to emulate the polycrystalline microstructure of an additively manufactured single-phase (austenitic) high-Mn steel serving as the model material which was meticulously experimentally characterized in previous studies [1,2]. Moreover, using the experimental characterization results of the model material, the adapted physics-based crystal plasticity constitutive model was developed, validated, and calibrated within the context of the same preceding studies.

To uncover the role of grain size anisotropy in the anisotropic mechanical response of polycrystals, a numerical approach is essential. Firstly, conducting experimental uniaxial deformation tests on the same material along numerous different axes presents significant challenges, if not impossibilities. Secondly, it is nearly impossible to experimentally separate the morphological texture from the crystallographic texture, as these two microstructural textures are intricately interconnected. In other words, it is highly improbable to experimentally obtain a polycrystal with a strongly anisotropic grain size (i.e., a strong morphological texture) while simultaneously having random distributions of crystallographic orientation and misorientation (i.e., lacking crystallographic texture). In practice, the processing conditions that produce strongly anisotropic grain morphologies (e.g., additive manufacturing), also give rise to strong crystallographic textures. Both experimental observations and numerical simulations confirm that columnar grains in AM metals arise due to steep thermal gradients along the build direction (BD) and high cooling rates during metal AM, especially in powder bed fusion, resulting in directional solidification and epitaxial grain growth that frequently aligns closely with <001> ∥ BD [3–10]. Consequently, AM metals tend to exhibit a preferential crystallographic texture corresponding to <001> ∥ BD.

The adapted physics-based crystal plasticity constitutive model does not account for the following advanced features: strain gradient, gradients of micro-state variables, dislocation density fluxes, and slip transmission across grain boundaries. These advanced constitutive modeling features can be incorporated in physics-based constitutive models for full-field crystal plasticity simulations of the deformation of polycrystals that solely exhibit dislocation slip [95,96]. However, conducting large-strain full-field crystal plasticity simulations using a highly nonlinear physics-based constitutive model that simultaneously accounts for dislocation slip, twinning, and the aforementioned modeling features is prohibitively expensive. These constitutive modeling features could provide a more realistic representation of grain boundary hardening mechanisms, potentially amplifying the size effect on strain hardening. Therefore, as a future prospect, we recommend conducting a similar study on another additively manufactured single-phase polycrystal as the model material, where dislocation slip serves as the sole crystal plasticity mechanism, while adopting a physics-based constitutive model that encompasses the aforementioned advanced features. This can be achieved by adopting a physics-based constitutive model that encompasses the aforementioned advanced features. By doing so, one can ensure a more accurate consideration of enhanced strain hardening near grain boundaries.

If a polycrystalline aggregate is perfectly textureless in terms of both crystallography and grain morphology, its macroscopic mechanical response will be fully isotropic regardless of the adopted crystal plasticity constitutive model and its associated parameters. Nevertheless, the crystal plasticity constitutive model and its parameters have a strong impact on the microscopic/constitutive mechanical response of intragranular material points. However, in the presence of any crystallographic and/or morphological textures, the degree of anisotropy in the macroscopic mechanical response may be influenced by the crystal plasticity constitutive model and parameters. Therefore, it is important to investigate the impact of constitutive parameters, such as initial dislocation density, on the anisotropic grain size effect.

## 6. Conclusions

We have formally introduced the concept of axial grain size from which we derived mean axial grain size, effective grain size, and grain size anisotropy as robust morphological descriptors capable of effectively representing highly complex grain morphologies. A discrete sample of single-phase polycrystalline aggregate was instantiated as an RVE. The distributions of crystallographic orientation and misorientation in the generated RVE are nearly random. However, the RVE does incorporate the typical morphological features found in additively manufactured microstructures, such as distinct grain size heterogeneity and anisotropic grain size owing to its pronounced columnar grain morphology. Therefore, any observable anisotropy in the macroscopic mechanical response of the instantiated sample is primarily a result of its anisotropic grain size. We conducted meso-scale full-field crystal plasticity simulations to analyze the mechanical responses of the RVE under uniaxial tension along various axes. These simulations employed a high-fidelity physics-based constitutive model, which was developed, calibrated, and validated in prior studies. Subsequently, we estimated the grain size and strain hardening anisotropies associated with the RVE. The main findings of this study are summarized as follows:

- To establish robust and efficient experimental or computational process-structure-properties linkages, it is crucial to represent the microstructure sufficiently and concisely. The meso-scale microstructure (i.e., mesostructure) of polycrystalline materials comprises two equally important aspects: grain morphology and crystallography. The conventional morphological descriptors, which rely on idealized grain shapes, are generally unsuitable for representing the complex and anisotropic grain morphologies in polycrystalline aggregates, including those typically found in additively manufactured metals.

- Considering the importance of the grain size effect on the mechanical response of polycrystals, it is sensible to represent the state of polycrystalline grain morphology using a single-variable morphological descriptor related to grain boundary spacing, namely axial grain size. Grain size anisotropy and effective grain size, which are rigorously defined based on the notion of axial grain size, offer a concise and effective means of representing complex and anisotropic grain morphology states in polycrystals.

- The advantages of the introduced descriptors over the conventional morphological descriptors mainly lie in their ability to effectively capture the collective grain size anisotropy (i.e., collective anisotropy of grain boundary spacing) and the effective grain size of the microstructure. The limitations of the conventional morphological descriptors are related to their multitude leading to loss of correlations, their inadequacy to characterize the collective grain size anisotropy of the microstructure, and the assumption of idealized shapes which are insufficient to represent anomalous and anisotropic grain shapes commonly observed in additively manufactured microstructures.

- Uniaxial strain hardening is the most appropriate measure for representing anisotropy in the macroscopic mechanical response of polycrystals, particularly that related to grain size anisotropy. Mean uniaxial strain hardening and strain hardening anisotropy, defined based on the uniaxial strain hardening response, can effectively characterize the anisotropy in the mechanical response of polycrystals.

- In a crystallographically textureless single-phase polycrystal, a relatively small (mean axial) grain size along a certain axis leads to a relatively high (mean) uniaxial strain hardening along that axis, and vice versa. The contribution of grain size anisotropy to the strain hardening anisotropy of polycrystalline aggregates can be described by the following inverse square relation:

$$\widehat{H}(\widehat{\boldsymbol{r}}) = k\left(\widehat{D}(\widehat{\boldsymbol{r}})\right)^{-2} + h;$$

where $\widehat{\boldsymbol{r}}$ represents an axis; $k$ and $h$ are two dimensionless constants; $\widehat{D}(\widehat{\boldsymbol{r}})$ denotes the grain size anisotropy function which is defined as the mean axial grain size associated with the axis $\widehat{\boldsymbol{r}}$ normalized by the effective grain size; and $\widehat{H}(\widehat{\boldsymbol{r}})$ denotes the strain hardening anisotropy function defined as the mean uniaxial strain hardening associated with the axis $\widehat{\boldsymbol{r}}$ normalized by the effective uniaxial strain hardening.

**Acknowledgments**

The authors acknowledge the support of the German Federal Ministry of Education and Research within the NanoMatFutur project "MatAM - Design of additively manufactured high-performance alloys for automotive applications" (Project ID: 03XP0264).

**Supplementary materials**

The Supplementary materials associated with this article, including the MATLAB script for axial grain size measurements and calculation of grain size anisotropy, can be found, in the GitHub repository at https://github.com/sahmotaman/Grain-Size-Anisotropy.

**Appendix A. Conventional grain morphology representation by idealized shapes**

The equivalent grain size of grain $g_i$ ($D_{eq}(g_i)$) within the microstructure domain $\Omega$ ($g_i \subseteq \Omega$) is defined as the diameter of equivalent sphere of grain $g_i$, which has the same volume as grain $g_i$: $v(g_i) \equiv \frac{\pi}{6}\left(D_{eq}(g_i)\right)^3$, where $v(g_i)$ denotes the volume of grain $g_i$. Therefore, the equivalent grain size of grain $g_i$ is defined as $D_{eq}(g_i) \equiv \sqrt[3]{\frac{6}{\pi}v(g_i)}$. Accordingly, the grain size distribution in the microstructure $\Omega$ can be represented by number- or volume-weighted probability density of the equivalent grain size ($P(D_{eq})$). Note that, in volume-weighted equivalent grain size distribution, the contribution of equivalent grain size of each grain to the equivalent grain size distribution of the aggregate is considered to be proportional to the volume of the grain.

Conventionally, to represent the grain morphological anisotropy and orientation the ellipsoid idealization is used, in which an equivalent ellipsoid is fitted to an arbitrarily shaped grain (best-fit ellipsoid) based on its zeroth-, first-, and second-order geometrical moments [11–14]. Each ellipsoid can be described by three semi-axes with the lengths $a$, $b$, and $c$ (so that $a \geq b \geq c$), aligned with the orthonormal basis $\{\hat{e}_a, \hat{e}_b, \hat{e}_c\}$. The morphological anisotropy can then be represented by two shape parameters associated with its equivalent/best-fit ellipsoid: equivalent minimum aspect ratio ($m \equiv \frac{c}{a}$) and equivalent maximum aspect ratio ($M \equiv \frac{b}{a}$), where $0 < m \leq M \leq 1$. Therefore, the lower the grain equivalent (ellipsoidal) aspect ratios, the higher the grain elongation and hence the grain morphological anisotropy. In addition, the grain morphological orientation can be represented by the spatial orientation of the grain equivalent (ellipsoidal) axes (GEA) with the (auxiliary) orthonormal basis $\{\hat{e}_a, \hat{e}_b, \hat{e}_c\}$ with respect to the standard/specimen Cartesian coordinate system $x_1 x_2 x_3$ having the orthonormal basis $\{\hat{e}_1, \hat{e}_2, \hat{e}_3\}$ using three Bunge-Euler angles $\boldsymbol{\Phi} \equiv (\varphi_1, \phi, \varphi_2)$ (Appendix D).

After fitting equivalent ellipsoids to the grains of the microstructure, the morphological anisotropy distribution using the ellipsoid model can be represented by number- or volume-weighted probability distribution of grain equivalent minimum and maximum aspect ratios. Furthermore, the morphological orientation distribution and thus the morphological texture can be represented by distribution of the orientation of GEA using the grain equivalent axes orientation distribution function (GEA-ODF). Given the microstructure domain $\Omega$, the GEA-ODF $P(\boldsymbol{\Phi})$ is the probability density of measuring the grain equivalent axes orientation $\boldsymbol{\Phi} \in SO(3)$ at a random point $\boldsymbol{x} \in \Omega$. GEA-ODF can be represented using series expansion of symmetrized generalized spherical harmonics [97,98] corresponding to the fundamental region of the orientation space $SO(3)$ associated with grain equivalent axes orientation. Grain equivalent axes orientation has the orthorhombic (local) symmetry. It should be noted that GEA-ODF does not contain any information regarding the distribution of morphological anisotropy and size. Moreover, the GEA-ODF can be plotted as pole figures with respect to the equivalent ellipsoidal semi-axes.

**Appendix B. Spherical harmonics representation and approximation**

For any two complex functions $f(\hat{\boldsymbol{r}})$ and $g(\hat{\boldsymbol{r}})$ on the surface of unit sphere ($S^2$), we define the following inner product:

$$\forall f, g : S^2 \to \mathbb{C}: \langle f | g \rangle = \int_{S^2} f^*(\hat{\boldsymbol{r}}) g(\hat{\boldsymbol{r}}) \mathrm{d}^2 \hat{\boldsymbol{r}}; \quad \hat{\boldsymbol{r}} \equiv (\theta, \varphi) \in S^2; \quad \mathrm{d}^2 \hat{\boldsymbol{r}} \equiv \sin\theta \, \mathrm{d}\theta \mathrm{d}\varphi; \tag{B.1}$$

where $\langle f | g \rangle \equiv \langle f | g \rangle^*$ denotes the inner product of $f$ with $g$; asterik $*$ indicates complex conjugate; $\hat{\boldsymbol{r}}$ denotes a point on the surface of unit sphere; $\theta$ and $\varphi$ are, respectively, the polar and azimuthal angles (spherical coordinates) of $\hat{\boldsymbol{r}}$; and $\mathrm{d}^2 \hat{\boldsymbol{r}}$ represents the invariant measure in $S^2$ (differential area on the surface of unit sphere in the vicinity of $\hat{\boldsymbol{r}}$). As a result, functions $f$ and $g$ belong to an infinite-dimensional Hilbert space having the inner product defined by Eq. (B.1). Using the inner product definition in Eq. (B.1), the $L^2$-norm for this Hilbert space is defined as:

$$\forall f: S^2 \to \mathbb{C}: \langle f \rangle \equiv \langle f | f \rangle \equiv \int_{S^2} f^*(\hat{\boldsymbol{r}}) f(\hat{\boldsymbol{r}}) \mathrm{d}^2 \hat{\boldsymbol{r}}; \tag{B.2}$$

where the functional $\langle f \rangle \in \mathbb{R}$ denotes the $L^2$-norm of $f$. The complex functions on the surface of unit sphere whose norm is finite (i.e., square-integrable functions on $S^2$) form a Hilbert space denoted by $L^2(S^2)$:

$$L^2(S^2) \equiv \{ f \mid f: S^2 \to \mathbb{C}, \langle f \rangle < \infty \}. \tag{B.3}$$

Two functions $f, g \in L^2(S^2)$, where $f \neq g$, are orthogonal and normal if and only if $\langle f|g \rangle = 0$, and $\langle f \rangle = \langle g \rangle = 1$, respectively.

The (surface) spherical harmonics functions (SHFs) are defined as follows:

$$Y_l^m(\hat{r}) \equiv Y_l^m(\theta, \varphi) \equiv \frac{1}{\sqrt{2\pi}} P_l^m(\cos\theta) e^{im\varphi}; \quad l \in \mathbb{N}_0; \quad m \in \{m \in \mathbb{Z} \mid |m| \leq l\};$$

$$P_l^m(z) \equiv \sqrt{\frac{(2l+1)}{2}} \sqrt{\frac{(l-m)!}{(l+m)!}} \frac{(-1)^m}{2^l l!} (1-z^2)^{\frac{m}{2}} \frac{d^{l+m}}{dz^{l+m}} (z^2-1)^l; \quad |z| \leq 1;$$

(B.4)

where $Y_l^m(\theta,\varphi): S^2 \to \mathbb{C}$ denotes the SHF of degree $l$ and order $m$; and $P_l^m(z): [-1,1] \to \mathbb{R}$ denotes the normalized associated Legendre polynomial of degree $l$ and order $m$. It can be shown that the SHFs are mutually orthonormal:

$$\forall Y_l^m, Y_{l'}^{m'}: \langle Y_l^m | Y_{l'}^{m'} \rangle = \delta_{mm'} \delta_{ll'} \quad \Rightarrow \quad \langle Y_l^m \rangle = \langle Y_{l'}^{m'} \rangle = 1;$$

(B.5)

where $\delta_{ij}$ is the Kronecker delta. The SHFs form a complete set of orthonormal basis functions for the Hilbert space of square-integrable functions on the surface of unit sphere ($L^2(S^2)$), and thus any square-integrable functions on the surface of unit sphere can be expanded as linear combination of SHFs:

$$\forall f \in L^2(S^2): f(\hat{r}) = \sum_{l=0}^{\infty} \sum_{m=-l}^{l} \tilde{f}_l^m Y_l^m(\hat{r}); \quad \tilde{f}_l^m \in \mathbb{C};$$

(B.6)

where $\tilde{f}_l^m$ is the Fourier coefficient of SHF of degree $l$ and order $m$ associated with the function $f$. Since the SHFs form a complete set of orthonormal basis functions for the Hilbert space $L^2(S^2)$, the Fourier coefficients $\tilde{f}_l^m$ are the inner product of the SHF $Y_l^m$ with the function $f \in L^2(S^2)$:

$$\tilde{f}_l^m \equiv \langle Y_l^m | f \rangle \equiv \int_{S^2} f(\hat{r}) \left(Y_l^m(\hat{r})\right)^* d^2\hat{r}.$$

(B.7)

The functional of Eq. (B.7) is known as (continuous) spherical Fourier transform.

Real square-integrable functions on the surface of unit sphere belong to the Hilbert space $L^2_\mathbb{R}(S^2) \equiv \{f \in L^2(S^2) \mid f: S^2 \to \mathbb{R}\}$ which is a subspace of $L^2(S^2)$ ($L^2_\mathbb{R}(S^2) \subset L^2(S^2)$). Therefore, any function $f \in L^2_\mathbb{R}(S^2)$ can be expressed as a linear combination of SHFs according to Eq. (B.6). The fulfillment of the following constraint on the Fourier coefficients of SHFs ensures that $f \in L^2_\mathbb{R}(S^2)$ is real-valued ($\forall \hat{r} \in S^2: f(\hat{r}) = f^*(\hat{r})$):

$$\forall f \in L^2_\mathbb{R}(S^2): \tilde{f}_l^{-m} = (-1)^m \left(\tilde{f}_l^m\right)^*.$$

(B.8)

The SHFs have definite parity that is they are either even or odd with respect to inversion about the origin:

$$Y_l^m(-\hat{r}) = (-1)^l Y_l^m(\hat{r}) \quad \Leftrightarrow \quad Y_l^m(\pi-\theta, \pi-\varphi) = (-1)^l Y_l^m(\theta, \varphi).$$

(B.9)

Moreover, real square-integrable functions with antipodal symmetry (i.e., point reflection symmetry with respect to origin) on the surface of unit sphere form the Hilbert space $L^2_{\mathbb{R}\pm}(S^2) \equiv \{f \in L^2_\mathbb{R}(S^2) \mid f(-\hat{r}) = f(\hat{r}), \hat{r} \in S^2\}$ which is a subspace of $L^2_\mathbb{R}(S^2)$ ($L^2_{\mathbb{R}\pm}(S^2) \subset L^2_\mathbb{R}(S^2)$), where the subscript $\pm$ denotes the antipodal symmetry. Since SHFs are mutually orthonormal:

$$\forall f \in L^2_{\mathbb{R}\pm}(S^2): Y_l^m(-\hat{r}) = Y_l^m(\hat{r}). \tag{B.10}$$

Combining Eqs. (B.9) and (B.10) leads to:

$$\forall f \in L^2_{\mathbb{R}\pm}(S^2), \forall l \in \{2n - 1 \mid n \in \mathbb{N}\}: \tilde{f}_l^m = 0. \tag{B.11}$$

Therefore, given Eqs. (B.6), (B.8), and (B.11), for any real square-integrable function $f$ with antipodal symmetry on the surface of unit sphere, the spherical harmonics representation reads:

$$\forall f \in L^2_{\mathbb{R}\pm}(S^2): f(\hat{r}) = \sum_{l=0,2,4}^{\infty} \sum_{m=0}^{l} \left( \tilde{f}_l^m Y_l^m(\hat{r}) + (-1)^{-m} \left( \tilde{f}_l^m \right)^* Y_l^{-m}(\hat{r}) \right). \tag{B.12}$$

As mentioned, any function $f \in L^2(S^2)$ can be represented as a linear combination of SHFs as they form a complete set of orthonormal bases for the infinite-dimensional Hilbert space $L^2(S^2)$. Nevertheless, since it is not feasible to calculate the infinite sum of Eqs. (B.6) or (B.12), function $f \in L^2(S^2)$ can be approximated by finite series expansion up to the SHF of finite degree $K$, known as (harmonic) bandwidth or cut-off degree:

$$\forall f \in L^2(S^2): f(\hat{r}) \cong {}^K f(\hat{r}) \equiv \sum_{l=0}^{K} \sum_{m=-l}^{l} \tilde{f}_l^m Y_l^m(\hat{r}); \quad K \in \mathbb{N}_0; \quad l \in \{l \in \mathbb{N}_0 \mid l \leq K\}; \tag{B.13}$$

where ${}^K f(\hat{r})$ is the approximant of $f$ with the bandwidth $K$ so that $\lim_{K \to \infty} {}^K f(\hat{r}) = f(\hat{r})$. The truncated/band-limited series of Eq. (B.13) has $(K + 1)^2$ terms, as $\forall l > K: \tilde{f}_l^m = 0$. Hence, increasing $K$ has a significant effect on the computation cost of the summation, as direct algorithms for such computation have the computational complexity of $\mathcal{O}(K^2)$. The bandwidth $K$ also plays an important role in the accuracy of the approximation of $f(\hat{r})$ by its truncated approximant as well as localization error. Therefore, the optimal choice of the bandwidth intricately depends on the desired degree of approximation and computational resources.

$f(\hat{r}) \in L^2(S^2)$ is often unknown a priori while it is measured (with some uncertainty) at a finite set of measurement/sample nodes. Let $\underline{f}$ be the measurement operator corresponding to function $f$, and $\boldsymbol{R} \equiv (\hat{r}_n)_{n=1}^N \in (S^2)^N$ an array of finite discrete measurement nodes; where $\hat{r}_n \equiv (\theta_n, \varphi_n) \in S^2; n \in \mathbb{N}, n \leq N; N \in \mathbb{N}$ denotes the number of nodes; and $(\hat{r}_n)_{n=1}^N \equiv (\hat{r}_1, \hat{r}_2, \ldots, \hat{r}_N)^T$ is the notation for $N$-tuple. Therefore, measuring $f$ at the measurement nodes gives the array of measurement values $\underline{\boldsymbol{F}} \equiv \underline{\boldsymbol{F}}(\boldsymbol{R}) \equiv \left( \underline{f}_n \right)_{n=1}^N$, where $\underline{f}_n \equiv \underline{f}(\hat{r}_n)$ denotes the measured value of $f$ at $\hat{r}_n$. The measurement nodes $(\hat{r}_n)_{n=1}^N$ are generally non-equispaced. In other words, the geodesic distances of pairs of nodes are not constant (i.e., the nodes are not uniformly distributed over the surface of the unit sphere). The objective is to determine the expansion coefficients $\tilde{f}_l^m$ of in Eq. (B.13) such that the distance between the measurements $\underline{\boldsymbol{F}}$ and the truncated estimates ${}^K \boldsymbol{F} \equiv {}^K \boldsymbol{F}(\boldsymbol{R}) \equiv \left( {}^K f_n \right)_{n=1}^N$ (where ${}^K f_n \equiv {}^K f(\hat{r}_n)$) is minimized. This distance is defined as the following norm of the weighted residuals [99]:

$$\eta \equiv \|\boldsymbol{W} \circ (\underline{\boldsymbol{F}} - {}^{K}\boldsymbol{F})\| \equiv \sum_{n=1}^{N} \left| w_n \left( \underline{f_n} - {}^{K}f_n \right) \right|^2 ; \tag{B.14}$$

where $\boldsymbol{W} \equiv \boldsymbol{W}(\boldsymbol{R}) \equiv (w_n)_{n=1}^{N}$ denotes the array of quadrature weights corresponding to nodes array $\boldsymbol{R}$; $w_n \equiv w(\hat{\boldsymbol{r}}_n) \geq 0$ denotes the non-negative quadrature weight associated with the measurement node $\hat{\boldsymbol{r}}_n$; and $\circ$ is the elementwise/Hadamard product operator. Note that $\boldsymbol{W}(\boldsymbol{R}): (S^2)^N \to (\mathbb{R}^+)^N$ denotes a quadrature rule on $S^2$. The expansion coefficients $\tilde{f}_l^m$ (up to the degree $K$) that minimize the norm $\eta$ can be sought by the least squares method, which is contingent on the evaluation of the following non-uniform/non-equispaced discrete spherical Fourier transform:

$$\tilde{f}_l^m \cong \sum_{n=1}^{N} w_n \, {}^{K}f(\hat{\boldsymbol{r}}_n) \left( Y_l^m(\hat{\boldsymbol{r}}_n) \right)^* . \tag{B.15}$$

The numerical integration of Eq. (B.15) corresponding to the quadrature rule $\boldsymbol{W}(\boldsymbol{R})$ is an approximation of the definite integral of Eq. (B.7). A suitable quadrature rule for numerical integration on the unit sphere is based on spherical Voronoi partition/tessellation. The Voronoi partition on $S^2$ associated with $N \in \mathbb{N}$ scattered nodes making the discrete node set $\mathrm{R} \equiv \{\hat{\boldsymbol{r}}_n \in S^2 | n \leq N, n \in \mathbb{N}\}$ reads:

$$\omega(\mathrm{R}) \equiv \{\omega(\mathrm{R}, \hat{\boldsymbol{r}}_n) \subset S^2 | \hat{\boldsymbol{r}}_n \in \mathrm{R}\}; \quad S^2 = \bigcup_{n=1}^{N} \omega(\mathrm{R}, \hat{\boldsymbol{r}}_n);$$

$$\omega(\mathrm{R}, \hat{\boldsymbol{r}}_n) \equiv \{\hat{\boldsymbol{r}} \in S^2 | d_{S^2}(\hat{\boldsymbol{r}}, \hat{\boldsymbol{r}}_n) \leq d_{S^2}(\hat{\boldsymbol{r}}, \hat{\boldsymbol{r}}_k), \forall \hat{\boldsymbol{r}}_k \in (\mathrm{R} - \{\hat{\boldsymbol{r}}_n\})\}; \quad \hat{\boldsymbol{r}}_n \in \mathrm{R}; \tag{B.16}$$

where $\omega(\mathrm{R}, \hat{\boldsymbol{r}}_n)$ denotes the (spherical) Voronoi cell corresponding to the node $\hat{\boldsymbol{r}}_n$; $\omega(\mathrm{R})$ represents the spherical Voronoi partition associated with the node set R; and $d_{S^2}(\hat{\boldsymbol{r}}, \hat{\boldsymbol{r}}_n)$ denotes the geodesic distance (i.e., length of the shortest path) between the points $\hat{\boldsymbol{r}}$ and $\hat{\boldsymbol{r}}_n$ on the surface of the unit sphere. Based on the preceding definitions, we define the following Voronoi indictor/characteristic functions and Voronoi weights:

$$w_n \equiv w(\mathrm{R}, \hat{\boldsymbol{r}}_n) \equiv \int_{S^2} \chi_n(\hat{\boldsymbol{r}}) \mathrm{d}^2 \hat{\boldsymbol{r}} \equiv \int_{\omega(\mathrm{R}, \hat{\boldsymbol{r}}_n)} \mathrm{d}^2 \hat{\boldsymbol{r}} ; \quad \sum_{n=1}^{N} w_n = 4\pi; \quad \hat{\boldsymbol{r}}_n \in \mathrm{R};$$

$$\chi_n(\hat{\boldsymbol{r}}) \equiv \begin{cases} 1 & : \hat{\boldsymbol{r}} \in \omega(\mathrm{R}, \hat{\boldsymbol{r}}_n) \\ 0 & : \hat{\boldsymbol{r}} \notin \omega(\mathrm{R}, \hat{\boldsymbol{r}}_n) \end{cases} ; \tag{B.17}$$

where $w_n \equiv w(\mathrm{R}, \hat{\boldsymbol{r}}_n)$ and $\chi_n(\hat{\boldsymbol{r}})$ denote the Voronoi weight and Voronoi indictor function associated with the node $\hat{\boldsymbol{r}}_n \in \mathrm{R}$, respectively. The array of Voronoi weights $\boldsymbol{W} \equiv \boldsymbol{W}(\boldsymbol{R}) \equiv (w_n)_{n=1}^{N}$ defines the quadrature rule used in Eqs. (B.14) and (B.15). Direct algorithms to compute the Voronoi partition on $S^2$ and the corresponding array of Voronoi quadrature weights $\boldsymbol{W}$ have quadratic complexity in the number of nodes ($\mathcal{O}(N^2)$). However, the algorithm proposed and implemented by Renka [100] computes the Voronoi cells and their associated weights for $N$ nodes on $S^2$ with $\mathcal{O}(N \log N)$ arithmetical operations.

Least squares estimation of expansion coefficients $\tilde{f}_l^m$ through minimizing the norm $\eta$ (Eq. (B.14)) by any iterative algorithm requires evaluation of the summations of Eqs. (B.13) and (B.15) for many times. Direct algorithms for calculation of the array of expansion coefficients $\tilde{f}_l^m$ by Eq. (B.15) and the array of band-limited function estimates ${}^{K}f(\hat{\boldsymbol{r}}_n)$ both have the complexity of $\mathcal{O}(NK^2)$. However, fast algorithms to compute the aforementioned arrays with complexity of $\mathcal{O}(K \log^2 K + N)$ [101,102], which are based on the non-equispaced fast Fourier transform (NFFT) algorithm [103,104] and implemented in MATLAB®-based MTEX toolbox [88], can be used in combination with the least squares algorithm to achieve higher computational efficiency.

**Appendix C. Uniaxial strain hardening**

Let the deformation of the initial/undeformed (continuum) microstructure domain/configuration $\Omega_0 \subset \mathbb{R}^3$ (at time $t = t_0$) to the current microstructure domain $\Omega \subset \mathbb{R}^3$ (at time $t > t_0$) be defined by the smooth time-dependent map $\chi: \Omega_0 \to \Omega$, so that $\boldsymbol{x} \equiv \chi(\boldsymbol{x}_0, t)$ where $\boldsymbol{x}$ denotes the current position of the material particle ($\boldsymbol{x} \in \Omega$) with the initial position $\boldsymbol{x}_0 \equiv \chi^{-1}(\boldsymbol{x}, t) \in \Omega_0$. Accordingly, the deformation gradient tensor (**F**) is defined as $\mathbf{F}(\boldsymbol{x}, t) \equiv \nabla_0 \boldsymbol{x}$, where $\nabla_0 \equiv \frac{\partial}{\partial \boldsymbol{x}_0}$ denotes the gradient operator with respect to the undeformed domain. Let mixed velocity and traction boundary conditions be imposed on the disjoint parts of $\partial \Omega$ ($\partial \Omega$ denotes the boundary of the microstructure domain $\Omega$ at time $t$). The balance of linear momentum for $\Omega$ in the absence of considerable body and inertia forces can be expressed by $\forall \boldsymbol{x} \in \Omega$: $\nabla_0 \cdot \mathbf{P} = \mathbf{0}$, where $\mathbf{P}(\boldsymbol{x}, t)$ denotes the first Piola-Kirchhoff stress tensor which is the work conjugate of **F**. The deformation rate is related to stress rate through the underlying constitutive relationship $\dot{\mathbf{P}} \equiv \mathbb{L} : \dot{\mathbf{F}}$, where the fourth-order tensor $\mathbb{L} \equiv \frac{\partial \mathbf{P}}{\partial \mathbf{F}}$ denotes the elasto-viscoplastic tangent stiffness modulus which is a spatiotemporal function ($\mathbb{L} \equiv \mathbb{L}(\boldsymbol{x}, t)$).

The uniaxial strain hardening is associated with the macroscopic mechanical response when the microstructure is subjected to homogenous uniaxial (tensile or compressive) deformation along a specific axis at constant strain rate and temperature. Homogenous macroscopic boundary conditions demand that the macro-mechanical quantities (macroscopic deformation rate and stress on $\partial \Omega$) are the volume averages of the corresponding micromechanical fields (microscopic deformation rate and stress in $\Omega$):

$$\overline{\dot{\mathbf{F}}}(t) = \frac{1}{V} \int_\Omega \dot{\mathbf{F}}(\boldsymbol{x}, t) \mathrm{d}^3 \boldsymbol{x}; \quad \overline{\mathbf{P}}(t) = \frac{1}{V} \int_\Omega \mathbf{P}(\boldsymbol{x}, t) \mathrm{d}^3 \boldsymbol{x}; \tag{C.1}$$

where $\overline{\dot{\mathbf{F}}}(t)$ and $\overline{\mathbf{P}}(t)$ denote the macroscopic/homogenized rate of deformation gradient and first Piola-Kirchhof stress tensors. The macroscopic mechanical boundary conditions associated with homogenous uniaxial deformation can be imposed by prescribing the components of $\overline{\dot{\mathbf{F}}}(t)$ and $\overline{\mathbf{P}}(t)$ in a mutually exclusive way (with respect to a single reference frame).

Let the reference Cartesian coordinate system $x_1 x_2 x_3$ with the orthonormal basis $\{\hat{\boldsymbol{e}}_1, \hat{\boldsymbol{e}}_2, \hat{\boldsymbol{e}}_3\}$ be transformed to the auxiliary frame $x_1' x_2' x_3'$ with the orthonormal basis $\{\hat{\boldsymbol{e}}_1', \hat{\boldsymbol{e}}_2', \hat{\boldsymbol{e}}_3'\}$ by a passive rotation so that the axis $\hat{\boldsymbol{r}}$ be parallel to the $x_3'$ axis ($\hat{\boldsymbol{r}}(\theta, \varphi) = \hat{\boldsymbol{e}}_3'$). This transformation can be accomplished using Bunge-Euler angles (Appendix D) so that $\varphi_1 = \varphi + \frac{\pi}{2}$, $\phi = \theta$, and $\varphi_2 = -\frac{\pi}{2}$, which corresponds to the orthogonal rotation/transformation matrix $\mathbf{Q}(\varphi_1, \phi, \varphi_2) = \mathbf{Q}\left(\varphi + \frac{\pi}{2}, \theta, -\frac{\pi}{2}\right)$. Consequently, $\mathbf{Q}\hat{\boldsymbol{r}} = \hat{\boldsymbol{e}}_3'$. Consider the microstructure $\Omega$ subjected to the constant homogenous macroscopic uniaxial nominal/engineering strain rate $\dot{\varepsilon}_0$ along the axis $\hat{\boldsymbol{r}}$ at constant temperature $T$. Note that the $\dot{\varepsilon}_0 > 0$ and $\dot{\varepsilon}_0 < 0$ correspond to uniaxial tension and compression, respectively. The constant mixed boundary conditions associated with the displacement-controlled uniaxial deformation along the axis $\hat{\boldsymbol{r}}$ can be enforced by prescribing the complementary components of $\overline{\dot{\mathbf{F}}}$ and $\overline{\mathbf{P}}$ with respect to the frame $x_1' x_2' x_3'$ as follows:

$$\overline{\dot{F}}'_{33} = \dot{\varepsilon}_0; \quad \overline{\dot{F}}'_{12} = \overline{\dot{F}}'_{21} = \overline{\dot{F}}'_{13} = \overline{\dot{F}}'_{31} = \overline{\dot{F}}'_{23} = \overline{\dot{F}}'_{32} = 0; \quad \overline{P}'_{11} = \overline{P}'_{22} = 0; \tag{C.2}$$

where $\dot{\bar{F}}'_{ij} \equiv \dot{\bar{\mathbf{F}}}: \hat{e}'_i \otimes \hat{e}'_j$ and $\bar{P}'_{ij} \equiv \bar{\mathbf{P}}: \hat{e}'_i \otimes \hat{e}'_j$, respectively, denote the components of the second-order tensors $\dot{\bar{\mathbf{F}}}$ and $\bar{\mathbf{P}}$ in the auxiliary frame $x'_1 x'_2 x'_3$, whereas $\dot{\bar{F}}_{ij} \equiv \dot{\bar{\mathbf{F}}}: \hat{e}_i \otimes \hat{e}_j$ and $\bar{P}_{ij} \equiv \bar{\mathbf{P}}: \hat{e}_i \otimes \hat{e}_j$, respectively, denote the components of $\dot{\bar{\mathbf{F}}}$ and $\bar{\mathbf{P}}$ in the standard reference frame $x_1 x_2 x_3$. Consequently, the components of $\dot{\bar{\mathbf{F}}}$ and $\bar{\mathbf{P}}$ with respect to the frames $x_1 x_2 x_3$ and $x'_1 x'_2 x'_3$ are related by the following transformations:

$$\dot{\bar{F}}_{ij} = Q^T_{ik} \dot{\bar{F}}'_{kl} Q_{lj}; \quad \dot{\bar{\mathbf{F}}} \equiv \dot{\bar{F}}_{ij} \hat{e}_i \otimes \hat{e}_j \equiv \dot{\bar{F}}'_{ij} \hat{e}'_i \otimes \hat{e}'_j;$$
$$\bar{P}_{ij} = Q^T_{ik} \bar{P}'_{kl} Q_{lj}; \quad \bar{\mathbf{P}} \equiv \bar{P}_{ij} \hat{e}_i \otimes \hat{e}_j \equiv \bar{P}'_{ij} \hat{e}'_i \otimes \hat{e}'_j; \tag{C.3}$$

where $Q_{ij} \equiv \hat{e}_i \cdot \hat{e}'_j$ and $Q^T_{ij} \equiv Q_{ji}$ denote the components of the rotation matrix $\mathbf{Q}$ and its transpose $\mathbf{Q}^T$, respectively; and implicit summation is assumed over the repeated indices.

Based on the polar decomposition theorem, the non-singular deformation gradient ($\mathbf{F}$) is uniquely decomposed into the product of the symmetric left stretch tensor ($\mathbf{V}$) and the orthogonal rotation tensor ($\mathbf{R}$):

$$\mathbf{F} \equiv \mathbf{VR}; \quad \mathbf{V} \equiv \mathbf{V}^T; \quad \mathbf{R}^{-1} \equiv \mathbf{R}^T. \tag{C.4}$$

The true work-conjugate pairs, the true/Cauchy stress tensor ($\boldsymbol{\sigma}(\mathbf{x}, t)$) and the true/logarithmic strain tensor ($\boldsymbol{\varepsilon}(\mathbf{x}, t)$), at point $\mathbf{x} \in \Omega$ and time $t$ are related to $\mathbf{F}(\mathbf{x}, t)$ and $\mathbf{P}(\mathbf{x}, t)$ as follows:

$$\boldsymbol{\sigma} \equiv \frac{1}{J} \mathbf{PF}^T; \quad J \equiv \det \mathbf{F};$$
$$\boldsymbol{\varepsilon} \equiv \ln(\mathbf{V}) = \ln(\lambda_i) \mathbf{v}_i \otimes \mathbf{v}_i; \tag{C.5}$$

where $J$ is the Jacobian of the deformation map; and $\lambda_i$ and $\mathbf{v}_i$ denote the eigenvalues and eigenvectors of $\mathbf{V}$, respectively. As a result of the homogenous macroscopic boundary conditions:

$$\bar{\boldsymbol{\varepsilon}}(t) = \frac{1}{V} \int_\Omega \boldsymbol{\varepsilon}(\mathbf{x}, t) \mathrm{d}^3 \mathbf{x}; \quad \bar{\boldsymbol{\sigma}}(t) = \frac{1}{V} \int_\Omega \boldsymbol{\sigma}(\mathbf{x}, t) \mathrm{d}^3 \mathbf{x}; \tag{C.6}$$

where $\bar{\boldsymbol{\varepsilon}}(t)$ and $\bar{\boldsymbol{\sigma}}(t)$ are the homogenous macroscopic true strain and stress tensors (at $\partial\Omega$), respectively. The true stress and strain tensors are decomposed to their corresponding deviatoric, and volumetric/hydrostatic parts as follows:

$$\bar{\boldsymbol{\varepsilon}}^d \equiv \bar{\boldsymbol{\varepsilon}} - \bar{\boldsymbol{\varepsilon}}^v; \quad \bar{\boldsymbol{\varepsilon}}^v \equiv \frac{1}{3} (\mathbf{I}: \bar{\boldsymbol{\varepsilon}}) \mathbf{I};$$
$$\bar{\boldsymbol{\sigma}}^d \equiv \bar{\boldsymbol{\sigma}} - \bar{\boldsymbol{\sigma}}^h; \quad \bar{\boldsymbol{\sigma}}^h \equiv \frac{1}{3} (\mathbf{I}: \bar{\boldsymbol{\sigma}}) \mathbf{I}; \tag{C.7}$$

where $\bar{\boldsymbol{\varepsilon}}^d$ and $\bar{\boldsymbol{\varepsilon}}^v$ denote the deviatoric and volumetric splits of the macroscopic true strain tensor $\bar{\boldsymbol{\varepsilon}}$, respectively; $\bar{\boldsymbol{\sigma}}^d$ and $\bar{\boldsymbol{\sigma}}^h$ stand for the deviatoric and hydrostatic splits of the macroscopic true stress tensor $\bar{\boldsymbol{\sigma}}$, respectively; $\mathbf{I}$ is the second-order unit/identity tensor. The equivalent/von-Mises macroscopic stress and strain (scalars) are defined as:

$$\varepsilon \equiv \sqrt{\frac{2}{3}} \, \|\bar{\boldsymbol{\varepsilon}}^d\|; \quad \sigma \equiv \sqrt{\frac{3}{2}} \, \|\bar{\boldsymbol{\sigma}}^d\|; \tag{C.8}$$

where $\varepsilon$ and $\sigma$ represent the equivalent macroscopic true stress and strain, respectively; and $\|\mathbf{T}\| \equiv \sqrt{\mathbf{T}\colon\mathbf{T}}$ denotes the 2-norm of second-order tensor $\mathbf{T}$.

The macroscopic strain hardening is defined as the derivative of deviatoric macroscopic true stress with respect to deviatoric macroscopic true strain: $\bar{\mathbb{H}} \equiv \frac{\partial \bar{\boldsymbol{\sigma}}^d}{\partial \bar{\boldsymbol{\varepsilon}}^d}$, where $\bar{\mathbb{H}}$ denotes the macroscopic strain hardening (fourth-order) tensor at time $t$. Note that the microscopic/local elasto-viscoplastic tangent stiffness modulus tensor $\mathbb{L} \equiv \frac{\partial \mathbf{P}}{\partial \mathbf{F}}$, which represents the underlying constitutive response, is closely related to the microscopic strain hardening tensor $\mathbb{H} \equiv \frac{\partial \boldsymbol{\sigma}^d}{\partial \boldsymbol{\varepsilon}^d}$. Moreover, the instantaneous equivalent macroscopic strain hardening or tangent modulus ($H$) is defined as:

$$H \equiv \frac{d\sigma}{d\varepsilon} \equiv \frac{\dot{\sigma}}{\dot{\varepsilon}}. \tag{C.9}$$

**Appendix D. Orientation representation by Bunge-Euler angles**

The relative orientation of two Cartesian reference frame belongs to the 3D rotation group SO(3) which is the group of all rotations about the origin of Euclidean 3-space. Such an orientation can be described by a number of orientation representations, such as Euler angles, orientation matrix, and the so-called neo-Eulerian (axis-angle, Rodrigues–Frank, conformal, homochoric, and unit quaternion) vectors [105–108]. Each representation has distinct advantages and disadvantages with respect to computational efficiency and data visualization and interpretation. Nonetheless, in the crystalline materials science, the Euler angles representation is the most commonly used orientation representation. Eulerian parameterization is particularly convenient to represent orientation distribution functions (ODFs) as linear combination of the generalized spherical harmonics [97,98], but they do not necessarily convey the most efficient visualization and interpretation relative to the neo-Eulerian representations [89,109]. Recently, efficient algorithms have been developed for fast computation of ODFs in terms of Euler angles based on kernel density estimation [86,87]. Nevertheless, ODF representation as series expansion over the hyperspherical harmonics in terms of quaternions is also a viable alternative to the standard Eulerian approach for ODF [110,111].

In the orientation representation by Euler angles, an arbitrary orientation of an auxiliary frame relative to the standard frame is described by three consecutive *elemental rotations* about predefined axes. The name Euler angles is reserved for those decompositions for which two of the three axes are equal, as in $x_3 x_1 x_3$ or $x_3 x_2 x_3$. Hence, there are six possible Euler angle conventions. In Materials Science, the most used definition of Euler angles is the Bunge convention, which has an axis triplet $x_3 x_1 x_3$, with the corresponding rotation angles triplet $(\varphi_1, \phi, \varphi_2)$ (applied from left to right) [97,98].

Formally, Euler angles are used to describe a second-order (passive) rotation/transformation matrix $\mathbf{Q}$ from a standard (right-handed) Cartesian frame $x_1 x_2 x_3$ with the orthonormal basis $\{\hat{\boldsymbol{e}}_1, \hat{\boldsymbol{e}}_2, \hat{\boldsymbol{e}}_3\}$ to an auxiliary (right-handed) Cartesian frame $x'_1 x'_2 x'_3$ having the orthonormal basis $\{\hat{\boldsymbol{e}}'_1, \hat{\boldsymbol{e}}'_2, \hat{\boldsymbol{e}}'_3\}$, so that $Q_{ij} \equiv \hat{\boldsymbol{e}}_i \cdot \hat{\boldsymbol{e}}'_j$ being the components of the rotation matrix $\mathbf{Q}$. This transformation is defined by a non-unique sequence of three elemental rotations. In the Bunge notation, the coordinate transformation using Bunge-Euler angles $(\varphi_1, \phi, \varphi_2)$ includes the following successive passive elemental rotations [97,98]:

(1) Rotation of the standard basis $\{\hat{e}_1, \hat{e}_2, \hat{e}_3\}$ about $\hat{e}_3$ by the angle $\varphi_1$ ($0 \leq \varphi_1 \leq 2\pi$) resulting in $\{\hat{e}_1^{(1)}, \hat{e}_2^{(1)}, \hat{e}_3\}$, which is represented by:

$$\mathbf{Q}_3(\varphi_1) \equiv \begin{bmatrix} \cos\varphi_1 & \sin\varphi_1 & 0 \\ -\sin\varphi_1 & \cos\varphi_1 & 0 \\ 0 & 0 & 1 \end{bmatrix}. \tag{D.1}$$

(2) Rotation of $\{\hat{e}_1^{(1)}, \hat{e}_2^{(1)}, \hat{e}_3\}$ basis about $\hat{e}_1^{(1)}$ by the angle $\phi$ ($0 \leq \phi \leq \pi$) giving $\{\hat{e}_1^{(1)}, \hat{e}_2^{(2)}, \hat{e}_3^{(1)}\}$, which is expressed as:

$$\mathbf{Q}_1(\phi) \equiv \begin{bmatrix} 1 & 0 & 0 \\ 0 & \cos\phi & \sin\phi \\ 0 & -\sin\phi & \cos\phi \end{bmatrix}. \tag{D.2}$$

(3) Rotation of $\{\hat{e}_1^{(1)}, \hat{e}_2^{(2)}, \hat{e}_3^{(1)}\}$ basis about $\hat{e}_3^{(1)}$ by the angle $\varphi_2$ ($0 \leq \varphi_2 \leq 2\pi$) yielding $\{\hat{e}_1^{(2)}, \hat{e}_2^{(3)}, \hat{e}_3^{(1)}\}$ which is aligned with $\{\hat{e}_1', \hat{e}_2', \hat{e}_3'\}$ and described by:

$$\mathbf{Q}_3(\varphi_2) \equiv \begin{bmatrix} \cos\varphi_2 & \sin\varphi_2 & 0 \\ -\sin\varphi_2 & \cos\varphi_2 & 0 \\ 0 & 0 & 1 \end{bmatrix}. \tag{D.3}$$

Therefore,

$$v_i' = Q_{ij}v_j; \quad T_{ij}' = Q_{ik}T_{kl}Q_{lj}^T; \quad \mathbf{Q}(\varphi_1, \phi, \varphi_2) \equiv \mathbf{Q}_3(\varphi_2)\mathbf{Q}_1(\phi)\mathbf{Q}_3(\varphi_1); \tag{D.4}$$

where implicit summation is assumed over the repeated indices; $\mathbf{Q}(\varphi_1, \phi, \varphi_2) \in SO(3)$ is the orthogonal rotation/orientation/transformation matrix associated with the Bunge-Euler angles $(\varphi_1, \phi, \varphi_2)$ ($\mathbf{Q}^T = \mathbf{Q}^{-1}$); superscript $T$ indicates transpose; $Q_{ij}$ and $Q_{ij}^T \equiv Q_{ji}$ are the components of the rotation matrix $\mathbf{Q}$ and its transpose $\mathbf{Q}^T$, respectively; $v_j$ and $v_i'$ denote the components of vector $\mathbf{v}$ with respect to the reference frames $x_1x_2x_3$ and $x_1'x_2'x_3'$, respectively: $\mathbf{v} \equiv v_j\hat{e}_j \equiv v_i'\hat{e}_i'$; and $T_{kl}$ and $T_{ij}'$ denote the components of the second-order tensor $\mathbf{T}$ with respect to the frames $x_1x_2x_3$ and $x_1'x_2'x_3'$, respectively: $\mathbf{T} \equiv T_{kl}\hat{e}_k \otimes \hat{e}_l \equiv T_{ij}'\hat{e}_i' \otimes \hat{e}_j'$.

It should be noted that it is necessary to use a region of the Bunge-Euler space containing each physically distinct orientation precisely once. It can be shown that $(\varphi_1, \phi, \varphi_2)$ and $(\varphi_1 + \pi, 2\pi - \phi, \varphi_2 + \pi)$ produce the exact same transformation matrix $\mathbf{Q}$. Therefore, the fundamental/asymmetrical region/zone of Bunge-Euler space is defined by $\{(\varphi_1, \phi, \varphi_2) \mid 0 \leq \varphi_1 < 2\pi; \ 0 \leq \phi \leq \pi; \ 0 \leq \varphi_2 < 2\pi\}$ which is the fundamental region without considering any additional symmetries (triclinic symmetry) in either the standard frame or the auxiliary frame. Considering symmetries associated with the standard frame and/or the auxiliary frame significantly reduces the size of the fundamental region.

## Appendix E. Crystal plasticity constitutive model

For the crystal constitutive relationship $\dot{\mathbf{P}} \equiv \mathbb{L}:\dot{\mathbf{F}}$ (Appendix C), we adapted the physics-based crystal plasticity constitutive model developed by Motaman et al. [1]. The physics-based crystal plasticity constitutive model explicitly accounts for dislocation slip and twinning as plastic deformation mechanisms at the associated deformation (slip/twin) systems. A model based on the continuum theory of crystal finite strain is used to project the deformation gradient and stress tensor at material points on the deformation systems in terms of (resolved) plastic shear strain rates and shear stresses. The mechanical response at the underlying deformation systems of each material point is (incrementally and implicitly) evaluated using physically motivated constitutive formulations. The resulting sets of nonlinear partial differential equations are then integrated (using the general Euler method with adaptive damping factor), linearized (using the tangent-based iterative Newton-Raphson scheme) and solved via a self-consistent semi-implicit return mapping (predictor-corrector) algorithm implemented in the modular crystal plasticity code DAMASK [91].

*E.1. Kinematics*

At single-crystal material points, the (total) deformation gradient tensor ($\mathbf{F}$) is multiplicatively decomposed into its elastic ($\mathbf{F}_e$) and isochoric plastic ($\mathbf{F}_p$) splits [112]:

$$\mathbf{F} = \mathbf{F}_e \mathbf{F}_p; \quad J \equiv \det(\mathbf{F}) = \det(\mathbf{F}_e) > 0; \quad \det(\mathbf{F}_p) = 1; \tag{E.1}$$

where subscript 0 represents the initial/undeformed/reference configuration; subscripts $e$ and $p$ denote elastic and plastic, respectively; and $J$ is the Jacobian of the deformation map. The main postulate through the aforementioned elasto-plastic decomposition is that the crystal plastic deformation occurs by flow of material through the crystal lattice without distorting the lattice itself, while the material point undergoes lattice distortion (giving rise to lattice stresses) due to the elastic deformation gradient. In this picture, the hypothetical configuration defined by the transformation $\mathbf{F}_p$ is referred to as the plastic/relaxed/unloaded/intermediate configuration. Accordingly, the (total) velocity gradient tensor ($\mathbf{L}$) can be expressed as:

$$\mathbf{L} \equiv \frac{\partial \mathbf{v}}{\partial \mathbf{x}} = \dot{\mathbf{F}} \mathbf{F}^{-1} = \mathbf{L}_e + \mathbf{L}_p; \quad \mathbf{v} \equiv \frac{\partial \mathbf{x}}{\partial t}; \quad \mathbf{L}_e \equiv \dot{\mathbf{F}}_e \mathbf{F}_e^{-1}; \quad \mathbf{L}_p \equiv \mathbf{F}_e \underline{\mathbf{L}}_p \mathbf{F}_e^{-1}; \quad \underline{\mathbf{L}}_p \equiv \dot{\mathbf{F}}_p \mathbf{F}_p^{-1}; \tag{E.2}$$

where $\mathbf{v}$ is the velocity vector; $t$ is time; and the underscore indicates the tensor quantity in the plastic configuration. Furthermore, through the polar decomposition theorem, the non-singular elastic deformation gradient tensor is uniquely decomposed into the product of the orthogonal elastic rigid-body rotation tensor ($\mathbf{R}_e$) and the symmetric right elastic stretch tensor ($\mathbf{U}_e$):

$$\mathbf{F}_e = \mathbf{R}_e \mathbf{U}_e; \quad \mathbf{R}_e^{-1} = \mathbf{R}_e^{\mathrm{T}}; \quad \mathbf{U}_e = \mathbf{U}_e^{\mathrm{T}}; \quad \Rightarrow \quad \mathbf{U}_e^2 = \mathbf{C}_e \equiv \mathbf{F}_e^{\mathrm{T}} \mathbf{F}_e; \tag{E.3}$$

where $\mathbf{C}_e$ is referred to as the elastic right Cauchy-Green deformation tensor. Therefore, the lattice rotation from the plastic configuration can be represented by $\mathbf{R}_e^{\mathrm{T}}$:

$$\mathbf{Q}(\boldsymbol{\varphi}) = \mathbf{R}_e^{\mathrm{T}}; \quad \mathbf{Q}^{-1} = \mathbf{Q}^{\mathrm{T}}; \tag{E.4}$$

where $\mathbf{Q}$ is the orientation matrix as a function of three Bunge-Euler angles $\boldsymbol{\varphi} \equiv (\varphi_1, \phi, \varphi_2)$ (Appendix D) at the arbitrary time $t$. Moreover, the initial orientation of the crystal lattice basis with respect to the reference frame is accommodated by initialization of the plastic deformation gradient by a virtual (initial) deformation (pure rotation) step [113]:

$$\mathbf{F}_{p0} \equiv \mathbf{Q}_0 \equiv \mathbf{Q}(\boldsymbol{\varphi}_0); \quad \mathbf{F}_0 = \mathbf{F}_{e0} \mathbf{F}_{p0} = \mathbf{I}; \quad \Rightarrow \quad \mathbf{F}_{e0} = \mathbf{R}_{e0} \mathbf{U}_{e0} = \mathbf{R}_{e0} = \mathbf{Q}_0^{-1} = \mathbf{Q}_0^{\mathrm{T}}; \quad \mathbf{U}_{e0} \equiv \mathbf{I}; \tag{E.5}$$

where the index 0 represents the initial (virtual) step ($t = 0$); and $\boldsymbol{\varphi}_0 \equiv (\varphi_{10}, \phi_0, \varphi_{20})$ is the set of three Bunge-Euler angles describing the initial/undeformed crystallographic lattice orientation of the considered material point with respect to the frame of reference. The virtual deformation step by $\mathbf{F}_{p0} \equiv \mathbf{Q}_0$ guarantees that the lattice basis in the plastic configuration always coincides with the reference coordinate system.

*E.2. Crystal plasticity*

The plastic velocity gradient at the given material point in the plastic configuration is estimated as follows:

$$\underline{\mathbf{L}}_p \cong \sum_{\alpha=1}^{N_{sl}} \dot{\gamma}_{sl}^{\alpha} \underline{\mathbf{Z}}_{sl}^{\alpha} + \sum_{\beta=1}^{N_{tw}} \dot{\gamma}_{tw}^{\beta} \underline{\mathbf{Z}}_{tw}^{\beta};$$

$$\underline{\mathbf{Z}}_{sl}^{\alpha} = \mathbf{F}_e^{-1} \mathbf{Z}_{sl}^{\alpha} \mathbf{F}_e; \quad \underline{\mathbf{Z}}_{sl}^{\alpha} \equiv \hat{\underline{\mathbf{b}}}_{sl}^{\alpha} \otimes \hat{\underline{\mathbf{n}}}_{sl}^{\alpha}; \quad \mathbf{Z}_{sl}^{\alpha} \equiv \hat{\mathbf{b}}_{sl}^{\alpha} \otimes \hat{\mathbf{n}}_{sl}^{\alpha};$$

$$\underline{\mathbf{Z}}_{tw}^{\beta} = \mathbf{F}_e^{-1} \mathbf{Z}_{tw}^{\beta} \mathbf{F}_e; \quad \underline{\mathbf{Z}}_{tw}^{\beta} \equiv \hat{\underline{\mathbf{b}}}_{tw}^{\beta} \otimes \hat{\underline{\mathbf{n}}}_{tw}^{\beta}; \quad \mathbf{Z}_{tw}^{\beta} \equiv \hat{\mathbf{b}}_{tw}^{\beta} \otimes \hat{\mathbf{n}}_{tw}^{\beta}; \quad \text{(E.6)}$$

where subscripts sl and tw indicate that the corresponding quantities are associated with slip and twin, respectively; $N_{sl}$ is the number of slip systems; and $N_{tw}$ is the number of twin systems; superscripts $\alpha$ and $\beta$ are slip and twin systems indices, respectively; $\gamma$ is the plastic shear strain; $\mathbf{Z}$ is the Schmid tensor; $\boldsymbol{b}$ is the slip/twin Burgers vector; $\hat{\boldsymbol{b}}$ is the slip/twin Burgers direction (unit) vector (normalized slip/twin Burgers vector); $\hat{\boldsymbol{n}}$ is the slip/twin plane normal (unit) vector; and the sign $\hat{\ }$ indicates normalization ($\hat{\bullet} = \frac{\bullet}{|\bullet|}$). There are twelve slip and twin systems in fcc crystal ($N_{sl}^{fcc} = N_{tw}^{fcc} = 12$). Moreover, in fcc crystal symmetry, $\frac{a_0}{2}\langle 0\,1\,\bar{1}\rangle$ and $\frac{a_0}{2}\langle 1\,1\,\bar{2}\rangle$, respectively, represent slip Burgers vector ($\underline{\boldsymbol{b}}_{sl}^{\alpha}$) and twin Burgers vector ($\underline{\boldsymbol{b}}_{tw}^{\beta}$) families, while $\frac{1}{\sqrt{3}}\{1\,1\,1\}$ represents both the slip plane normal ($\hat{\underline{\boldsymbol{n}}}_{sl}^{\alpha}$) and twin plane normal ($\hat{\underline{\boldsymbol{n}}}_{tw}^{\beta}$) families. Note that Eq. (E.6) only holds through the following assumptions [114]: (i) twins can be sheared by dislocation slip in a compatible manner to the parent/surrounding matrix; and (ii) any potentially different evolution of slip resistance within them is negligible.

*E.3. Crystal elasticity*

The mean elastic stiffness tensor in the plastic configuration ($\bar{\underline{\mathbb{C}}}_e$) at a given material point is homogenized as follows to account for the contributions of twins and matrix [115,116]:

$$\bar{\underline{\mathbb{C}}}_e = (1-f_{tw})\,\underline{\mathbb{C}}_{e\,mt} + \sum_{\beta=1}^{N_{tw}} f_{tw}^{\beta}\,\underline{\mathbb{C}}_{e\,tw}^{\beta}; \quad f_{tw} = \sum_{\beta=1}^{N_{tw}} f_{tw}^{\beta};$$

$$\left[\underline{\mathbb{C}}_{e\,tw}^{\beta}\right]_{ijkl} = \left[\underline{\mathbb{C}}_{e\,mt}\right]_{pqrs} \left[\underline{\mathbf{T}}_{mt-tw}^{\beta}\right]_{ip} \left[\underline{\mathbf{T}}_{mt-tw}^{\beta}\right]_{jq} \left[\underline{\mathbf{T}}_{mt-tw}^{\beta}\right]_{kr} \left[\underline{\mathbf{T}}_{mt-tw}^{\beta}\right]_{ls}; \quad \underline{\mathbf{T}}_{tw}^{\beta} = 2\hat{\underline{\boldsymbol{n}}}_{tw0}^{\beta} \otimes \hat{\underline{\boldsymbol{n}}}_{tw0}^{\beta} - \mathbf{I}; \quad \text{(E.7)}$$

where the bar sign denotes mean/homogenization; $f$ represents volume fraction; $\mathbb{C}$ is the fourth-order stiffness tensor; subscript mt stands for matrix; $\underline{\mathbf{T}}_{mt-tw}^{\beta}$ represents the matrix transforming the lattice orientation in the parent matrix to the lattice orientation in the twinned region (twin system $\beta$) in the plastic configuration; and the square brackets [ ] is used to indicate index notation. Note that, in case of cubic (fcc/bcc) crystal symmetry, due to the symmetry of the matrix elastic stiffness tensor $\underline{\mathbb{C}}_{e\,mt}$, it only contains three independent components (namely the elastic constants $C_{11}$, $C_{12}$, and $C_{44}$) [117].

According to the general/three-dimensional Hook's law, the second Piola-Kirchoff (nominal) stress tensor in the plastic configuration ($\underline{\mathbf{S}}$) is calculated by (double) contraction (inner product) of $\bar{\underline{\mathbb{C}}}_e$ with its work conjugate pair, the elastic Green-Lagrange (nominal) strain tensor in the plastic configuration ($\underline{\mathbf{E}}_e$):

$$\underline{\mathbf{S}} = \bar{\underline{\mathbb{C}}}_e : \underline{\mathbf{E}}_e; \quad \underline{\mathbf{E}}_e = \frac{1}{2}(\mathbf{C}_e - \mathbf{I}). \quad \text{(E.8)}$$

Given Eq. (E.6) and considering the power conjugacy of the (asymmetric) Mandel stress (**M**) and velocity gradient (**L**) tensors [118]:

$$\dot{w}_p = \underline{\mathbf{M}} : \underline{\mathbf{L}}_p = \sum_{\alpha=1}^{N_{sl}} \tau_{sl}^{\alpha} \dot{\gamma}_{sl}^{\alpha} + \sum_{\beta=1}^{N_{tw}} \tau_{tw}^{\beta} \dot{\gamma}_{tw}^{\beta}; \quad \underline{\mathbf{M}} \equiv \underline{\mathbf{C}}_e \underline{\mathbf{S}};$$
$$\Rightarrow \quad \tau_{sl}^{\alpha} = \underline{\mathbf{M}} : \underline{\mathbf{Z}}_{sl}^{\alpha} \cong \underline{\mathbf{S}} : \underline{\mathbf{Z}}_{sl}^{\alpha}; \quad \tau_{tw}^{\beta} = \underline{\mathbf{M}} : \underline{\mathbf{Z}}_{tw}^{\beta} \cong \underline{\mathbf{S}} : \underline{\mathbf{Z}}_{tw}^{\beta}; \tag{E.9}$$

where $\dot{w}_p$ is the volumetric plastic deformation power at the considered material point; and $\tau$ is the resolved shear stress (RSS) at the corresponding deformation system. In general, the deformation of metallic materials are categorized as hypoelasto-viscoplastic. Hence, one typically assumes that elastic strains are very small compared to unity, so that $\mathbf{F}_e \approx \mathbf{I}$, leading to $\underline{\mathbf{C}}_e \approx \mathbf{I}$, which follows the useful approximation of $\underline{\mathbf{M}} \approx \underline{\mathbf{S}}$.

*E.4. Kinetics of slip*

In the framework of continuum dislocation dynamics, the notion of dislocation density as a micro-state variable is used to describe dislocation slip and its corresponding phenomena. The total dislocation density at slip system $\alpha$ is decomposed to unipolar and dipolar dislocation densities:

$$\rho_t^{\alpha} \equiv \rho_u^{\alpha} + \rho_d^{\alpha}; \quad \rho_x \equiv \sum_{\alpha=1}^{N_{sl}} \rho_x^{\alpha}; \quad x \in \{t, u, d\}; \tag{E.10}$$

where $\rho$ denotes dislocation density; subscripts $t$, $u$ and $d$ stand for total, unipolar and dipolar, respectively; and $\rho$ without slip system superscript ($\alpha$) is the sum of the corresponding dislocation density type ($x$) over all the slip systems at the considered material point.

Dislocations in dipolar configuration are considered immobile (zero slip velocity), while unipolar dislocations can be mobile and thus contribute to the plastic shear strain rate according to the Orowan equation [119]:

$$\dot{\gamma}_{sl}^{\alpha} = b_{sl} \rho_u^{\alpha} \bar{v}_{sl\,u}^{\alpha}; \quad \bar{v}_{sl\,u}^{\alpha} = v_{sl0} \exp\left(-\frac{\Delta G_{sl}^{\alpha}}{k_B T}\right) \text{sign}(\tau_{sl}^{\alpha}); \quad \Delta G_{sl}^{\alpha} = \Delta F_{sl}\left(1 - \left(\frac{\tau_{sl\,eff}^{\alpha}}{\tau_{sl0}}\right)^{p_{sl\,t}}\right)^{p_{sl\,b}};$$
$$\tau_{sl\,eff}^{\alpha} = \ll |\tau_{sl}^{\alpha}| - \tau_{sl\,cr}^{\alpha} \gg; \tag{E.11}$$

where $b_{sl}$ denotes the slip Burgers length (magnitude of Burgers vector at slip systems: $|\boldsymbol{b}_{sl}|$), which is given by $b_{sl} = \frac{\sqrt{2}}{2} a_0$ ($a_0$ is the lattice constant); $\Delta G_{sl}^{\alpha}$ is the average Gibbs free energy difference (activation enthalpy) for bypassing short-range obstacles by mobile dislocations at slip system $\alpha$; $k_B$ is the Boltzmann constant; $T$ is the absolute temperature; $v_{sl\,0}$ is the reference mean slip speed (magnitude of velocity vector) of unipolar dislocations (mean unipolar dislocation speed at high temperatures); $\Delta F_{sl}$ is the mean thermal activation energy (Helmholtz free energy) for slip without the aid of external stress; $\bar{v}_{sl\,u}^{\alpha}$ is the mean slip velocity of unipolar dislocations at slip system $\alpha$ (represented in the local frame of the slip system $\alpha$); $\tau_{sl0}$ is the maximum short-range slip resistance (in average), also known as Peierls stress or solid solution strength; $\tau_{sl\,eff}^{\alpha}$ is the effective/viscous/rate-dependent/thermal/friction/over shear stress at slip system $\alpha$; $p_{sl\,t}$ and $p_{sl\,b}$ are the constitutive exponents that respectively describe the shape of the top and bottom of the short-range obstacle force-distance profile and are constrained by: $0 < p_{sl\,t} \leq 1$; $1 \leq p_{sl\,b} \leq 2$; $\ll \gg$ are the Macaulay brackets:

$\ll \bullet \gg = \frac{1}{2}(|\bullet| + \bullet)$; and $\tau^\alpha_{\text{sl cr}}$ is the mean critical/athermal/rate-independent/internal/back (resolved) shear stress to activate the slip at system $\alpha$.

The critical shear stress at slip system $\alpha$ ($\tau^\alpha_{\text{sl cr}}$) is the minimum shear stress at slip system $\alpha$ (in average) that needs to be overcome for bowed-out mobile dislocations to slip. $\tau^\alpha_{\text{sl cr}}$ is a function of (total) dislocation density and calculated by the generalized Taylor equation, which accounts for the anisotropic interactions between slip systems (i.e., latent hardening) [120]:

$$\tau^\alpha_{\text{sl cr}} = \mu b_{\text{sl}} \sqrt{\sum_{\dot{\alpha}=1}^{N_{\text{sl}}} A^{\alpha\alpha'} \rho_t^{\alpha'}}; \tag{E.12}$$

where $A^{\alpha\alpha'}$ is the slip interaction strength matrix entry representing strength of interaction between slip systems $\alpha$ and $\alpha'$; and $\mu$ is the crystal shear modulus which, in case of cubic (fcc/bcc) crystal symmetry, can be calculated based on the Voigt homogenization/approximation scheme as a function of the elastic constants as follows [117]: $\mu = \frac{1}{5}(C_{11} - C_{12} + 3C_{44})$. In the fcc crystal symmetry, the interaction matrix $A^{\alpha\alpha'}$ has $(N_{\text{sl}}^{\text{fcc}})^2 = 144$ components. The number of distinct entries is divided by two due to the diagonal symmetry of the matrix ($A^{\alpha\alpha'} = A^{\alpha'\alpha}$) and the occurrence of four $\langle 111 \rangle$ axes with ternary symmetry further divides it by twelve. As such, there are only six independent coefficients which are associated with six distinct types of interactions [121]. Three of them account for forest interactions between non-coplanar slip systems, resulting in the formation of the orthogonal/Hirth lock, the sessile/Lomer–Cottrell lock, and the Glissile junction. There are two coplanar interactions for dislocations slipping on parallel slip planes with parallel Burgers vector: the dipolar/self-interaction; or unparallel coplanar Burgers vectors: the cross-coplanar interaction. Finally, the strongest interaction, the collinear interaction, occurs between dislocations slipping on two slip planes that are cross-slip planes with respect to each other.

The evolution of unipolar and dipolar dislocation densities at slip system $\alpha$ is given as follows [1,122]:

$$\dot{\rho}_u^\alpha = \left(\frac{1}{\Lambda_{\text{sl}}^\alpha} - 2h_{d\,\text{max}}^\alpha \rho_u^\alpha\right)\frac{|\dot{\gamma}_{\text{sl}}^\alpha|}{b_{\text{sl}}}; \quad \dot{\rho}_d^\alpha = 2(h_{d\,\text{max}}^\alpha \rho_u^\alpha - h_{d\,\text{min}} \rho_t^\alpha)\frac{|\dot{\gamma}_{\text{sl}}^\alpha|}{b_{\text{sl}}} - \frac{4|\bar{v}_{\text{cl}\,d}^\alpha|}{h_{d\,\text{max}}^\alpha - h_{d\,\text{min}}}\rho_d^\alpha; \tag{E.13}$$

where $\Lambda_{\text{sl}}^\alpha$ is the mean free path (MFP) for dislocation slip at system $\alpha$; $h_{d\,\text{min}}$ is the minimum height (slip plane spacing) of stable dislocation dipoles (in average); $h_{d\,\text{max}}^\alpha$ is the maximum height of stable dipoles at slip system $\alpha$ (in average); and $|\bar{v}_{\text{cl}\,d}^\alpha|$ is the average (out-of-plane) climb speed/rate of dipolar dislocations located at slip system $\alpha$.

The upper bound of the stable dipole height is calculated as follows [123]:

$$h_{d\,\text{max}}^\alpha = \frac{\mu b_{\text{sl}}}{8\pi(1-\nu)|\tau_{\text{sl}}^\alpha|}; \quad h_{d\,\text{min}} \leq h_{d\,\text{max}}^\alpha \leq \Lambda_{\text{sl}}^\alpha; \tag{E.14}$$

where $\nu$ is the Poisson's ratio which, in case of cubic (fcc/bcc) crystal symmetry, can be calculated based on the Voigt homogenization scheme as follows [117]: $\nu = \frac{C_{11}+4C_{12}-2C_{44}}{4C_{11}+6C_{12}+2C_{44}}$. The mean dislocation climb velocity is a function of the average dipole height and temperature [124]:

$$|\bar{v}_{\text{cl}\,d}^\alpha| \cong \frac{D_v V_{\text{cl}} \xi^\alpha}{b_{\text{sl}}^2 k_B T}; \quad D_v = D_{v0} \exp\left(-\frac{Q_v}{k_B T}\right); \quad \xi^\alpha = \frac{\mu b_{\text{sl}}^2}{2\pi(1-\nu)\bar{h}_d^\alpha}; \quad \bar{h}_d^\alpha = \frac{h_{d\,\text{min}} + h_{d\,\text{max}}^\alpha}{2}; \tag{E.15}$$

where $V_{cl}$ is the activation volume for climb; $\xi^\alpha$ is the average normal force (parallel to the slip plane normal) exerted over unit length of the dislocation line, under which dipole partners attract one another; $Q_v$ is the mean thermal activation energy for vacancy diffusion; $D_v$ is self-diffusivity or self/vacancy diffusion coefficient; $D_{v0}$ is reference/pre-exponential self-diffusivity; and $\bar{h}_d^\alpha$ is the average dipole height at slip system $\alpha$.

The MFP for slip at system $\alpha$ is known as the mean distance for slip of a mobile dislocation before its motion is impeded by an obstacle (dislocation accumulation/storage). However, more fundamentally, the slip MFP represents the mean (mobile) dislocation segment length, and thus governs the rate of dislocation multiplication. The slip MFP has confining contributions due to various existing sources of obstacles (grain boundaries, forest dislocations, and twin boundaries), which can be homogenized using the following harmonic mean/mixture law [34,125]:

$$\frac{1}{\Lambda_{sl}^\alpha} = \frac{1}{D_m(\widehat{\underline{b}}_{sl}^\alpha, \boldsymbol{x})} + \frac{1}{\lambda_f^\alpha} + \frac{1}{\lambda_{sl-tw}^\alpha}; \tag{E.16}$$

where $D_m(\widehat{\underline{b}}_{sl}^\alpha, \boldsymbol{x})$ the minimum distance to grain boundary along the axis $\hat{\boldsymbol{r}} = \widehat{\underline{b}}_{sl}^\alpha$ at the material point $\boldsymbol{x}$ (given by Eq. (10) in Section 2.3): $\lambda_f^\alpha$ is the effective forest dislocations spacing at slip system $\alpha$; and $\lambda_{sl-tw}^\alpha$ is the mean spacing among twin boundaries interacting with slipping mobile dislocation at slip system $\alpha$.

The most significant contribution to the slip MFP arises from the interaction of mobile dislocations with forest dislocations, which are the dislocations of other systems that pierce their slip plane. The well-known relationship for the effective forest spacing ($\lambda_f \propto \frac{1}{\sqrt{\rho_t}}$) can be generalized to account for the contribution of different slip systems ($\alpha'$) on the forest hardening of a specific slip system ($\alpha$) [126]:

$$\frac{1}{\lambda_f^\alpha} = \frac{1}{c_f}\sqrt{\sum_{\alpha'=1}^{N_{sl}} B_f^{\alpha\alpha'} \rho_t^{\alpha'}}; \quad B_f^{\alpha\alpha'} = |\widehat{\underline{n}}_{sl}^\alpha \cdot \widehat{\underline{t}}_e^{\alpha'}|; \quad \widehat{\underline{t}}_e^{\alpha'} = \widehat{\underline{n}}_{sl}^{\alpha'} \times \widehat{\underline{b}}_{sl}^{\alpha'}; \tag{E.17}$$

where $c_f \equiv c_{sl-sl}$ is the constitutive coefficient associated with the forest interactions; $B_f^{\alpha\alpha'}$ denotes the forest projection matrix, which is the matrix detailing the anisotropic forest interaction coefficients; and $\boldsymbol{t}_e$ is the edge dislocation (tangent) line vector.

In case of the occurrence of twinning, Fullman's stereological relationship [127] can be invoked to obtain an estimate for the mean twin boundary spacing. The generalized/modified form of Fullman's relationship, accounting for anisotropic interaction of different twin systems on restricting the MFP for slip at system $\alpha$ is given by [34]:

$$\frac{1}{\lambda_{sl-tw}^\alpha} = \frac{1}{c_{sl-tw}} \sum_{\beta=1}^{N_{tw}} B_{sl-tw}^{\alpha\beta} \frac{f_{tw}^\beta}{\delta_{tw}(1-f_{tw})}; \quad B_{sl-tw}^{\alpha\beta} = \begin{cases} 0 & : \widehat{\underline{n}}_{sl}^\alpha = \widehat{\underline{n}}_{tw}^\beta \\ 1 & : \widehat{\underline{n}}_{sl}^\alpha \neq \widehat{\underline{n}}_{tw}^\beta \end{cases}; \tag{E.18}$$

where $c_{sl-tw}$ is a constitutive coefficient associated with the effective topology of twins and its impact on the slip resistance; $\delta_{tw}$ represents the mean twin thickness; and $B_{sl-tw}^{\alpha\beta}$ denotes the anisotropic slip-twin interaction matrix, defined based on the assumption that only a twin system ($\beta$), which is non-coplanar/secant with a specific slip system ($\alpha$) confines its slip MFP.

*E.5. Kinetics of twinning*

The kinetics of twinning and the evolution of twin volume fraction read [34,128]:

$$\dot{\gamma}_{tw}^{\beta} = \gamma_{tw}\dot{f}_{tw}^{\beta}; \quad \dot{f}_{tw}^{\beta} = (1-f_{tw})V_{tw}^{\beta}\dot{n}_{tw}^{\beta}P_{tw}^{\beta}; \quad V_{tw}^{\beta} = \frac{\pi}{4}\left(\Lambda_{tw}^{\beta}\right)^2\delta_{tw}; \quad (E.19)$$

where $\gamma_{tw}$ is the characteristic plastic shear strain induced by twinning, which in case of fcc crystals is $\gamma_{tw}^{fcc} = \sqrt{2}/2$; $V_{tw}^{\beta}$ is the average volume of twins on twin system $\beta$; $\dot{n}_{tw}^{\beta}$ is the average number density of twin stacking faults nucleated per unit time at twin system $\beta$; $P_{tw}^{\beta}$ is the probability density for propagation/bowing-out of the corresponding twin nucleus to form a twin under the application of the homogenized RSS at twin system $\beta$ ($\tau_{tw}^{\beta}$); and $\Lambda_{tw}^{\beta}$ is the MFP for twinning at twin system $\beta$.

It is envisaged that the propagation of twins is mainly terminated at either grain boundaries or twin boundaries which are non-coplanar with the propagating twin [33]. As a result, the mean free path for twins at twin system $\beta$ is given by the following harmonic mean relationship [34]:

$$\frac{1}{\Lambda_{tw}^{\beta}} = \frac{1}{c_{tw-tw}}\left(\frac{1}{D\left(\hat{\underline{b}}_{tw}^{\beta},x\right)} + \frac{1}{\lambda_{tw-tw}^{\beta}}\right);$$

$$\frac{1}{\lambda_{tw-tw}^{\beta}} = \sum_{\beta'=1}^{N_{tw}} B_{tw-tw}^{\beta\beta'}\frac{f_{tw}^{\beta'}}{\delta_{tw}(1-f_{tw})}; \quad B_{tw-tw}^{\beta\beta'} = \begin{cases} 0 & :\hat{\underline{n}}_{tw}^{\beta} = \hat{\underline{n}}_{tw}^{\beta'} \\ 1 & :\hat{\underline{n}}_{tw}^{\beta} \neq \hat{\underline{n}}_{tw}^{\beta'} \end{cases}; \quad (E.20)$$

where $c_{tw-tw}$ is a constitutive parameter associated with the effective topology of twins and its impact on the strength of twin-twin interaction; $B_{tw-tw}^{\beta\beta'}$ is the anisotropic twin-twin interaction coefficients matrix, which is defined using the same rule as for $B_{sl-tw}^{\alpha\beta}$; and $D\left(\hat{\underline{b}}_{tw}^{\beta},x\right)$ denotes the grain boundary spacing (i.e. axial grain size) along the axis $\hat{r} = \hat{\underline{b}}_{sl}^{\alpha}$ at the material point $x$ (given by Eq. (9) in Section 2.3).

The twin propagation (from the existing twin stacking fault embryos) is a stochastic event, that requires sufficiently high local stress concentration, which can be maintained only if the mean RSS at the respective twin system is adequately high. We adopted the probabilistic treatment of the twin propagation process (based on its underlying physics) as a stochastic Poisson process described by the following cumulative probability density function:

$$P_{tw}^{\beta} = \exp\left(-\left(\frac{\ll\tau_{tw}^{\beta}\gg}{\tau_{tw\,cr}}\right)^{-p_{tw}}\right); \quad p_{tw} > 0; \quad (E.21)$$

where $p_{tw}$ is the constitutive exponent controlling the sigmoid shape of the associated cumulative probability density function $P_{tw}^{\beta}$, which is affected by the heterogeneity in elemental micro-segregation (distribution of alloying elements and its variance in lower scales); and $\tau_{tw\,cr}$ is the critical RSS (at twin systems) for twin propagation. $\tau_{tw\,cr}$ in fcc crystals based on twin nucleation model proposed by Mahajan and Chin [129], is given by [130]:

$$\tau_{tw\,cr} = \frac{\Gamma_{sf}}{3b_{tw}} + \frac{3\mu b_{tw}}{l_{tw}}. \quad (E.22)$$

where $\Gamma_{sf}$ is the stacking fault energy; $l_{tw}$ represents the mean length of twin nuclei; and $b_{tw}$ denotes the twin Burgers length (i.e., magnitude of Burgers vector at twin systems: $|\boldsymbol{b}_{tw}|$).

The nucleation of a twin stacking fault is a stochastic process, that requires sufficiently high local stress concentration, which can be maintained only if the mean RSS at the respective twin system is adequately high, while cross-slip (acting as another mechanism for relaxing the stress concentration) is an unfavorable competing mechanism [131–134]. Therefore, following Steinmetz et al. [130], we postulate the following relationship for $\dot{n}_{tw}^{\beta}$:

$$\dot{n}_{tw}^{\beta} = \dot{n}_{tw0}^{\beta} P_{sf}^{\beta}; \quad \dot{n}_{tw0}^{\beta} = \frac{1}{l_{tw}}\left(|\dot{\gamma}_{sl}^{\alpha}|\rho_t^{\alpha'} + |\dot{\gamma}_{sl}^{\alpha'}|\rho_t^{\alpha}\right);$$
$$\hat{\underline{n}}_{tw}^{\beta} = \hat{\underline{n}}_{sl}^{\alpha} = \hat{\underline{n}}_{sl}^{\alpha'}; \quad 3\underline{b}_{tw}^{\beta} = \underline{b}_{sl}^{\alpha} + \underline{b}_{sl}^{\alpha'}; \quad \alpha \neq \alpha';$$
(E.23)

where $P_{sf}^{\beta}$ is the probability density that cross-slip does not take place, which would allow a sufficient (local) dislocation accumulation rendering the stress concentration (at twin system $\beta$) necessary for dislocation dissociation/separation and the consequent stacking fault generation; and $\dot{n}_{tw0}^{\beta}$ is the (reference) average number density of potential twin stacking fault nuclei per unit time at twin system $\beta$. Eq. (E.23) is based on the twin nucleation mechanism in fcc crystals envisaged by Mahajan and Chin [129], in which two coplanar full $\frac{a_0}{2}\langle 01\bar{1}\rangle$ dislocations (where $a$ is the lattice constant) of different Burgers vectors $\underline{b}_{sl}^{\alpha} \neq \underline{b}_{sl}^{\alpha'}$ split into fault pairs and react on the primary slip plane to emanate three $\frac{a_0}{6}\langle 11\bar{2}\rangle$ Shockley partial dislocations with identical Burgers vector $\underline{b}_{tw}^{\beta}$ on successive parallel {111} planes. This results in an ordered three-layer stacking fault arrangement that can produce a twin.

Inspired by the relationship proposed by Kubin et al. [135] for the frequency of cross-slip occurrence, we suggest the following equation for $P_{sf}^{\beta}$:

$$P_{sf}^{\beta} = 1 - P_{cs}^{\beta}; \quad P_{cs}^{\beta} = \exp\left(-\frac{V_{cs}}{k_B T} \ll \tau_{sf\,cr} - \ll \tau_{tw}^{\beta} \gg\gg \right);$$
(E.24)

where $P_{cs}^{\beta}$ is the probability for cross-slip occurrence at slip systems conjugate (with parallel plane normal) to twin system $\beta$; $V_{cs}$ is the activation volume for cross-slip; and $\tau_{sf\,cr} > 0$ is the mean critical shear stress for dislocation dissociation. In fcc crystals, the critical shear stress to nucleate a three-layer stacking fault ($\tau_{sf\,cr}$), that can serve as a twin embryo, is given as a threshold for bringing the respective repulsive partial dislocations within the critical distance $r_c$ [129,136,137]:

$$\tau_{sf\,cr} = \frac{\mu b_{tw}}{2\pi}\left(\frac{1}{r_0 + r_c} + \frac{1}{2r_0}\right); \quad r_0 = \frac{(2+\nu)\mu b_{tw}^2}{8\pi(1-\nu)\Gamma_{sf}};$$
(E.25)

where $r_0$ is the equilibrium dissociation distance for Shockley partials in fcc crystals, in which the repulsive force due to the partial dislocations is balanced by the attractive force exerted by the stacking fault.

*E.6. Constitutive parameters*

The physics-based crystal plasticity constitutive model has been calibrated and validated for the model material (an additively manufactured austenitic high-Mn steel) in previous studies [1,2]. The values of the parameters of the physics-based crystal plasticity constitutive model, which depend on temperature ($T = 298$ K), as well as the crystal symmetry (fcc), the processing history, and the chemical composition of the model material (an additively manufactured austenitic high-Mn steel), are presented in Table E.1. The value of some of the listed constitutive parameters are directly adopted from the literature and the value of the rest of the parameters are determined through an iterative inverse calibration procedure to minimize the error between the simulated macroscopic stress-strain responses and their corresponding experimental counterparts.

**Table E.1.** Constitutive parameters of the physics-based crystal plasticity constitutive model.

| Category | Parameter | Symbol | Value | Unit | Reference |
|---|---|---|---|---|---|
| General | Lattice constant | $a_0$ | 0.361 | nm | [34,138,139] |
| | Elastic constants | $C_{11}$ | 175 | GPa | [140] |
| | | $C_{12}$ | 115 | | |
| | | $C_{44}$ | 135 | | |
| Initial micro-state variables | Initial unipolar dislocation density at each slip system | $\rho_{u0}^\alpha$ | $3 \times 10^{13}$ | m$^{-2}$ | |
| | Initial dipolar dislocation density at each slip system | $\rho_{d0}^\alpha$ | $10^{12}$ | m$^{-2}$ | [1] |
| | Initial twin volume fraction | $f_{tw0}$ | 0 | - | |
| Slip kinetics | Thermal activation (Helmholtz free) energy for slip | $\Delta F_{sl}$ | 0.9 | eV | [141] |
| | Reference slip speed of unipolar dislocations | $v_{sl0}$ | $5 \times 10^{-7}$ | m s$^{-1}$ | [141] |
| | Maximum short-range slip resistance (i.e., solid solution strength) | $\tau_{sl0}^\alpha$ | 44 | MPa | [1,141,142] |
| | Top shape parameter for short-range obstacle force-distance profile | $p_{sl\,t}$ | 1 | - | [1] |
| | Bottom shape parameter for short-range obstacle force-distance profile | $p_{sl\,b}$ | 1 | - | [1] |
| Recovery | Minimum stable dipole height | $h_{d\,min}$ | 6 | $b_{sl}$ | [122] |
| | Thermal activation energy for (climb) vacancy diffusion | $Q_v$ | 2.05 | eV | [1,136,143] |
| | Activation volume for climb | $V_{cl}$ | 1 | $b_{sl}^3$ | [122] |
| | Reference self-diffusion coefficient | $D_{v0}$ | 0.4 | cm$^2$ s$^{-1}$ | [142] |
| Dislocation junction/lock strengths in fcc crystals | Dipolar/self | $A_{dip}$ | 0.122 | - | |
| | Cross-coplanar | $A_{cop}$ | 0.122 | - | |
| | Collinear | $A_{col}$ | 0.625 | - | [121] |
| | Orthogonal/Hirth | $A_{ort}$ | 0.070 | - | |
| | Glissile | $A_{gls}$ | 0.137 | - | |
| | Sessile/Lomer-Cottrell | $A_{ses}$ | 0.122 | - | |
| Twinning kinetics | Stacking fault energy | $\Gamma_{sf}$ | 20.2 | mJ m$^{-2}$ | [1,144] |
| | Mean length of twin nuclei | $l_{tw}$ | 96 | nm | [1] |
| | Parameter controlling probability of twin formation | $p_{tw}$ | 3.5 | - | [1] |
| | Mean twin thickness | $\delta_{tw}$ | 0.2 | μm | [1] |
| | Activation volume for cross-slip | $V_{cs}$ | 15 | $b_{sl}^3$ | [145–147] |
| | Critical distance for proximity of repulsive partial dislocations to form twin's stacking fault nucleus | $r_c$ | 1 | nm | [147] |
| Mean free paths | Forest (slip-slip) interaction strength coefficient | $c_{sl-sl}$ | 16 | - | |
| | Slip-twin interaction strength coefficient | $c_{sl-tw}$ | 10 | - | [1] |
| | Twin-twin interaction strength coefficient | $c_{tw-tw}$ | 0.1 | - | |

## Appendix F. General grain size effect in polycrystals

The scale of the effective grain size marks three regimes of grain size effect on the macroscopic mechanical response of polycrystals, which are governed by domination of different micro-mechanisms of crystal plasticity:

1) Coarse-grained polycrystals and/or relatively high dislocation density: This regime is associated with grain boundary hardening (i.e., enhanced strain hardening near grain boundaries) stemming from (elasto-plastic) deformation incompatibilities at grain boundaries as well as grain boundary dislocation pileups and sources, which leads to increasing the macroscopic strain hardening and thus the initial/proof yield stress (i.e., stress at small strains) by decreasing the effective grain size [21,25,27,29–32,148–162].

2) Fine-grained polycrystals with relatively low dislocation density: In this regimes, the macroscopic mechanical response is associated with the appearance of yield point (or Lüders) phenomenon or discontinuous yielding that occurs owing to grain boundary strengthening, due to which by decreasing the effective grain size the lower yield point increases [16,56–65,156,163]. This grain size-dependent discontinuous yielding, which originates from inhomogeneous intergranular and intragranular elasto-plastic deformation, governs the early stage of deformation of fine-grained polycrystals with relatively low dislocation density and continues until the average intragranular shear resistance reaches the mean threshold for plasticity of grain boundaries (e.g., resistance for slip transmission across grain boundaries) [2].

3) Ultrafine-grained polycrystals (polynanocrystals or nanocrystals): This regime is associated with the grain size softening phenomenon or inverse Hall-Petch effect which corresponds to a decrease in the macroscopic yield stress by decreasing the effective grain size (i.e., negative Hall-Petch slope) [47–54]. This abnormal grain size dependence of mechanical response of ultrafine-grained polycrystals is due to the prevalence of stress-induced/shear-coupled grain boundary motion mechanisms (migration, sliding, and rotation) [40–46,55,164].

The grain size dependence of mechanical response of polycrystals in all of the above-mentioned regimes are often described by Hall-Petch-type relations. The classical Hall-Petch equation is an isotropic relation which describes the relationship between two scalar variables: the initial yield stress ($\sigma_y$) and the effective grain size in isotropic polycrystalline materials. The initial yield stress is taken as the *lower yield point* or *proof stress* (i.e., stress corresponding to a relatively small strain), respectively, where discontinuous and continuous yielding occurs. The Hall-Petch relation can be generalized to crystalline materials with anisotropic and textured microstructures to establish a link between the effective initial macroscopic yield stress $\bar{\bar{\sigma}}_y$ and the effective grain size $\bar{\bar{D}}$ as follows:

$$\bar{\bar{\sigma}}_y = \sigma_0 + k_{\mathrm{HP}} \bar{\bar{D}}^n; \qquad\qquad (\mathrm{F.1})$$

where $\sigma_0 > 0$ is a friction stress; and $k_{\mathrm{HP}} \in \mathbb{R}$ and $n < 0$ are known as the Hall-Petch coefficient/slope and scaling exponent, respectively. Note that, even though the Hall-Petch scaling exponent is typically taken as $n = -0.5$ with the support of a large body of data in the literature, the best value for the Hall-Petch exponent remains highly controversial and depending on individual datasets can vary in the range $-2 \leq n < 0$ [21,23,25,151,153–155,157,162,165–178].

In contrast to discontinuous yielding, there is no distinct initial yield point associated with a continuous yielding response. In the case of continuous yielding behavior, the transition from macroscopic fully elastic deformation to macroscopic elasto-plastic deformation (i.e., macroscopic plastic yielding) does not occur at an instant during deformation and is often associated with a notably wide strain range. Therefore, conventionally, the arbitrary notion of *proof stress* is used to represent the macroscopic initial yield stress in the presence of continuous yielding. Nonetheless, in order to describe the first regime of grain size effect (coarse-grained polycrystals with relatively high dislocation density and a continuous yielding response), it is more reasonable to use the effective strain hardening ($\bar{\bar{H}}$) instead of the initial yield stress ($\sigma_y$) since: (i) it does not require a distinct yield point and an arbitrary *proof stress*, (ii) the grain size effect in the first regime mainly originates from enhanced strain hardening due to grain boundary hardening, and (iii) since the strain hardening response is a derivative quantity (derivative of the stress response with respect to strain), it is more sensitive to microstructure and thus can better reveal the grain size effect.

Accordingly, the generalized Hall-Petch relation (Eq. (F.1)) can be rewritten in terms of the effective strain hardening $\bar{\bar{H}}$. Let $\varepsilon$ be the small (equivalent macroscopic) strain at which the effective initial macroscopic yield stress $\bar{\bar{\sigma}}_y$ is measured. Therefore, with assumption of continuous yielding, in the small strain interval $[0, \varepsilon]$, the effective stress and strain have an almost linear relationship. Hence, according to Eqs. (C.8), (C.9), (19), and (20), we can write $\bar{\bar{H}} \approx \frac{\bar{\bar{\sigma}}_y}{\varepsilon}$, where $\bar{\bar{H}}$ is close to but lower than the effective macroscopic elastic modulus of the material. Inserting it in Eq. (F.1) gives:

$$\bar{\bar{H}} = k_0 \bar{\bar{D}}^n + h_0; \quad k_0 \equiv \frac{k_{\mathrm{HP}}}{\varepsilon}; \quad h_0 \equiv \frac{\sigma_0}{\varepsilon}. \tag{F.2}$$

As mentioned, the first regime of grain size effect is characterized by domination of grain boundary hardening mechanisms including grain boundary dislocation pileups/sources and deformation incompatibilities at grain boundaries. As the effective grain size decreases, the grain boundary surface area density increases leading to a higher volumetric frequency of these grain boundary hardening mechanisms and thus a higher effective strain hardening. This is consistent with Eq. (F.2), where $n < 0$ and $k_0 > 0$.